\documentclass[preprint,pra,floatfix,amsmath,amssymb]{revtex4-2}
\usepackage{graphicx}
\usepackage{bm}
\usepackage{subcaption}
\usepackage{xcolor}
\usepackage{amsmath}
\usepackage{amssymb}

\usepackage{dcolumn}
\newcolumntype{d}[1]{D{.}{.}{#1}}
\newcolumntype{e}[1]{D{.}{}{#1}}

\usepackage{hyperref}
\hypersetup{colorlinks=true,citecolor=blue,linkcolor=blue,urlcolor=blue}

\begin{document}

\title{Thermophysical properties of argon gas from improved two-body interaction potential}

\author{Jakub Lang} \email{jakub.lang@chem.uw.edu.pl}
\author{Micha{\l} Przybytek}
\author{Micha{\l} Lesiuk}
\affiliation{Faculty of Chemistry, University of Warsaw, Pasteura 1, 02-093 Warsaw, Poland}


\begin{abstract}
\emph{Ab initio} interaction potential for the electronic ground state of the argon dimer has been developed. The potential uses previously calculated accurate Born--Oppenheimer interaction energies while significantly improving the description of relativistic effects by including the two-electron Darwin and orbit-orbit corrections. Moreover, leading-order quantum electrodynamics corrections to the potential are calculated and long-range retardation of the electromagnetic interactions is taken into account. Spectroscopic properties of the argon dimer are reported, such as the bond dissociation energy, positions of rovibrational levels, and rotational and centrifugal-distortion constants. Our potential supports eight rotationless vibrational states and the existence of a ninth level can neither be confirmed nor ruled out at the current accuracy level. Finally, thermophysical properties of the argon gas, including pressure and acoustic virial coefficients, as well as transport properties -- viscosity and thermal conductivity -- are evaluated using the developed potential. For the virial coefficients, the obtained \emph{ab initio} values are somewhat less accurate than the most recent experimental results. However, the opposite is true for the transport properties, where the theoretical results calculated in this work have significantly smaller uncertainties than the data derived from measurements.
\end{abstract}

\maketitle

\section{Introduction \label{sec:intro}}

Knowledge of accurate thermophysical properties of noble gases is critical in many areas of physics and chemistry and has been the subject of numerous experimental and theoretical studies. 
In particular, reliable values of thermophysical properties, such as the pressure, acoustic, dielectric, and refractivity virial coefficients, as well as transport properties, such as the viscosity and thermal conductivity, are needed in the field of metrology \cite{fellmuth2006determination,schmidt2007polarizability,moldover2014acoustic,gaiser2019highly,rourke2019refractive,gaiser2020primary,gaiser2022primary}. 
This is especially important for gas thermometry experiments, including the constant-volume gas thermometry \cite{fellmuth2006determination}, the acoustic gas thermometry \cite{moldover2014acoustic}, the dielectric-constant gas thermometry \cite{gaiser2019highly}, and the refractive-index gas thermometry \cite{rourke2019refractive}.

Many of the thermophysical properties of noble gases can be calculated theoretically using methods of statistical thermodynamics, provided that the potentials that take into account interactions in the system are known \cite{hirschfelder1964molecular,mcquarrie76}. 
In fact, computations for helium and neon gases show that the properties obtained from state-of-the-art \emph{ab initio} interaction potentials and interaction-induced polarizabilities have uncertainties similar to or even smaller than the best experimental data \cite{czachorowski2020,lang2023three,lang2023collision,hellmann2021thermophysicalNeon}. 
As such, they have been utilized for calibration of high-precision experimental equipment \cite{fellmuth2006determination,gaiser2019highly,gaiser2022primary}. 
Since argon represents an economical alternative to helium and neon, its thermophysical properties are of special interest. 
Their calculation from first principles necessitates the development of a reliable pair interaction potential, which is a challenging problem. 
Whereas for the lightest noble gas dimer, He$_2$, theoretical potentials have become more accurate than the empirical ones already in the mid-1990s \cite{Williams96,Korona97}, the empirical potentials for Ar$_2$, such as the one developed by Aziz \cite{aziz1993}, have generally been believed to be more reliable.
Until recently, the most accurate \emph{ab initio} argon dimer potentials have been developed by J\"ager \emph{et al.}\ \cite{jager2009} and Patkowski \emph{et al.}\ \cite{patkowski2005}, with the latter being later refined in Ref.~\cite{patkowski2010}. 
Other theoretical potentials published in recent years include the 2018 potential of Myatt \emph{et al.}\ \cite{myatt2018} based on recalibrated Morse/long-range model, the 2019 potential of Song and Yang \cite{song2019} with modified repulsive part of the Tang--Toennies model, simplified \textit{ab initio} atomic potential (SAAP) of Deiters and Sadus \cite{deiters2019two} from the same year, and the latest 2020 potential by Sheng \emph{et al.}\ \cite{sheng2020conformal}, where the conformality among two-body potentials of noble gas atoms is assumed.

It is worthwhile to compare spectroscopic parameters resulting from theoretical potentials with available experimental determinations. 
The empirical potential of Aziz \cite{aziz1993} predicts the minimum of the potential well as $-99.553$~cm$^{-1}$ at $R = 3.757$~\AA, while the \emph{ab initio} potentials of J\"ager \emph{et al.}\ \cite{jager2009} and Patkowski and Szalewicz \cite{patkowski2010} predict $-99.48$~cm$^{-1}$ and $-99.351$~cm$^{-1}$, respectively, both at $R = 3.762$~\AA. 
These results agree with the spectroscopic determination of Herman \emph{et al.}\ \cite{herman1988exp} which is $-99.2(10)$~cm$^{-1}$ at $R = 3.761(3)$~\AA. 
The most recent experimental determination of the ground-state rotational ($B_0$) and centrifugal-distortion ($D_0$) constants was performed by Mizuse \emph{et al.}\ \cite{mizuse2022} using time-resolved Coulomb explosion imagining. 
Based on spectroscopic analysis of the first 14 rotational levels, they obtained $B_0=0.057611(4)$~cm$^{-1}$ and $D_0=1.03(2)\times10^{-6}$~cm$^{-1}$. 
This can be compared to the theoretical results of J\"ager \emph{et al.}\ \cite{jager2009} who reported $B_0=0.05760$~cm$^{-1}$ without uncertainty estimates, and did not consider $D_0$. 
Theoretical values of $B_0 = 0.057589$~cm$^{-1}$ and $D_0 = 1.03\times10^{-6}$~cm$^{-1}$ resulting from the potential of Patkowski and Szalewicz \cite{patkowski2010} were calculated by Mizuse \emph{et al.}\ \cite{mizuse2022} using the energies of rotational levels computed in Ref.~\cite{sahraeian2019}. 

Recently, the question has emerged \cite{rivlin2019} of how many rotationless vibrational levels are supported by the electronic ground state of the argon dimer.
Sahraeian and Hadizadeh \cite{sahraeian2019} solved the momentum-space based Lippmann--Schwinger equation to obtain the vibrational states and reported nine bound levels for the potentials from Refs.~\cite{patkowski2005,patkowski2010}. 
This came after the 2016 study of Tennyson \emph{et al.}\ \cite{tennyson2016} based on the potential of Ref.~\cite{patkowski2010}, where eight bound vibrational levels were found using the $R$-matrix theory -- a weakly bound ninth state was assessed to be a numerical artifact. 
Eight bounds states were also obtained for the empirical potential of Myatt \emph{et al.}\ \cite{myatt2018} using {\sc LEVEL} program \cite{leroy2017level} to solve the nuclear Schr\"odinger equation. 
Somewhat later, Rivlin \emph{et al.}\ \cite{rivlin2019} published a study using refined $R$-matrix theory \cite{rivlin2019book} and also observed the ninth bound vibrational level for the potentials of Refs.~\cite{patkowski2005,patkowski2010}. 
Nevertheless, the existence of the ninth vibrational state and its possible energy is still unclear.
For example, Sahraeian and Hadizadeh \cite{sahraeian2019} predicted the energy of only $-0.20186\times10^{-6}$~cm$^{-1}$ for the potential from Ref.~\cite{patkowski2010}, while Rivlin \emph{et al.}\ \cite{rivlin2019} did not provide any specific value and concluded that the energy may be significantly different from the value obtained by Sahraeian and Hadizadeh. 

To meet the current requirements for the accuracy of theoretical predictions of thermophysical properties of noble gases, the \emph{ab initio} pair potentials used in the calculations have to account for effects beyond the non-relativistic electronic Schr\"odinger equation. 
In particular, the relativistic effects are expected to bring a considerable contribution to the interaction energy of the argon-argon dimer due to large nuclear charges of the atoms. 
The first calculation of the relativistic corrections to the two-body potential for this system were performed by Faas \emph{et al.}\ in 2000 \cite{faas2000}. 
They calculated the contribution of the relativistic effects near the minimum of the potential using the zeroth-order regular approximation (ZORA) to the Dirac equation \cite{Chang1986,vanLenthe1993,vanLenthe1994} and the second-order M\o{}ller--Plesset method for the treatment of correlation effects. 
More recent studies included relativistic corrections using other approaches. 
J\"ager \emph{et al.}\ \cite{jager2009} employed the Cowan--Griffin approximation within the first-order perturbation theory \cite{Cowan:76}, while Patkowski and Szalewicz \cite{patkowski2010} used the second-order Douglas--Kroll--Hess (DKH) relativistic Hamiltonian \cite{douglas1974,hess1986}. 
As noted in Ref.~\cite{patkowski2010}, a good agreement was observed between these two methods.
For example, in the region near the minimum of the potential the results differed by merely $0.012$~cm$^{-1}$ -- only about 2\% of the total relativistic correction. 
However, both approaches account only for the one-electron relativistic effects, while two-electron effects are completely neglected. 
These terms may have a substantial contribution to the final results.
As noted in Ref.~\cite{patkowski2010}, for the internuclear distance of $R = 3.75$~\AA, i.e., near the minimum of the potential, the two-electron terms amount to $-0.071$~cm$^{-1}$ or 11.6\% of the one-electron terms.  
Moreover, as the total two-electron relativistic correction vanishes as $R^{-4}$ with the increasing internuclear distance, i.e., more slowly than the one-electron correction vanishing as $R^{-6}$ \cite{Meath:66A}, its relative importance is likely to increase for larger distances.

In this work, we refine the two-body interaction potential of argon by the inclusion of the two-electron relativistic and leading-order quantum electrodynamics (QED) effects.
Moreover, the retardation effects are taken into account to properly describe the dissociation limit. 
The rovibrational levels supported by the potential are calculated to shed a light on the existence of the ninth bound vibrational state. 
Next, fully quantum calculations of the second pressure and acoustic virial coefficients of argon gas are performed. 
Transport properties of gaseous argon, such as viscosity and thermal conductivity, are also rigorously evaluated within the same framework.

Atomic units ($e=m_e=\hbar=4\pi\epsilon_0=1$, where $m_e$ is the electron mass) are used throughout the present work unless explicitly stated otherwise. The value of the electron spin $g$-factor is fixed and equal to $2$ and $\alpha = 1/137.035\,999\,084$ \cite{codata18} denotes the fine-structure constant.

\section{Pair potential for the argon dimer \label{sec:potential}}

\subsection{Born--Oppenheimer interaction energy \label{subsec:potBO}}

The data of Patkowski \emph{et al.}\ \cite{patkowski2005,patkowski2010} are currently the most accurate theoretical results for the non-relativistic interaction energy for two argon atoms. 
Their Born--Oppenheimer (BO) calculations are saturated with respect to both the basis set and electron excitation limits. 
Substantial improvements to this component of the two-body potential are not feasible for the foreseeable future with the current computational resources. 
For this reason, we reused the non-relativistic interaction energy calculated in Refs.~\cite{patkowski2005,patkowski2010} for 41 grid points. 

While the estimated uncertainties of the calculated BO energies for $R\ge2$~\AA{} are explicitly reported in Ref.~\cite{patkowski2010}, this is not the case for the data in the range $0.25\,\text{\AA}<R<2$~\AA{} from Ref.~\cite{patkowski2005}. 
Therefore, we assumed a conservative uncertainty estimation of 2\% for the latter dataset. 
We justify this choice by the observation that differences between extrapolated and not extrapolated results in Ref.~\cite{patkowski2005} are always less than 1.7\% for small distances. 

\subsection{Relativistic correction \label{subsec:potrel}}

For systems with nuclear charges below $Z\approx20$, the relativistic effects can be accounted for perturbatively as the expectation value of the Breit--Pauli Hamiltonian calculated with the non-relativistic wave function \cite{BeSal,Pachucki:04}. 
The Breit--Pauli Hamiltonian for closed-shell systems has the following form
\begin{equation} \label{rel:tot}
\hat{H}_\mathrm{BP}
=\hat{H}_\mathrm{mv}
+\hat{H}_\mathrm{D1}
+\hat{H}_\mathrm{D2}
+\hat{H}_\mathrm{oo}
+\hat{H}_\mathrm{ssc},
\end{equation}
where $\hat{H}_\mathrm{mv}$ is the mass-velocity operator,
\begin{equation} \label{rel:mv}
\hat{H}_\mathrm{mv} = 
-\frac{\alpha^2}{8} \sum_i \bm{p}_i^4,
\end{equation}
$\hat{H}_\mathrm{D1}$ and $\hat{H}_\mathrm{D2}$ are the one- and two-electron Darwin operators,
\begin{align} \label{rel:D1}
\hat{H}_\mathrm{D1} & 
= \alpha^2 \frac{\pi}{2} \sum_I\sum_i\,Z_I\,\delta(\bm{r}_{Ii}), 
\\ \label{rel:D2}
\hat{H}_\mathrm{D2} & 
= -\alpha^2 \pi \sum_{i<j} \delta(\bm{r}_{ij}),
\end{align}
$\hat{H}_\mathrm{oo}$ is the orbit-orbit operator,
\begin{equation} \label{rel:oo}
\hat{H}_\mathrm{oo} = -\frac{\alpha^2}{2} \sum_{i<j} \left[ 
 \frac{\bm{p}_i\cdot\bm{p}_j}{r_{ij}}
+\frac{\bm{r}_{ij}\cdot(\bm{r}_{ij}\cdot\bm{p}_j)\,\bm{p}_i}{r_{ij}^3}
\right],
\end{equation}
and $\hat{H}_\mathrm{ssc}$ is the spin-spin contact operator,
\begin{equation} \label{rel:ss}
\hat{H}_\mathrm{ssc}
= -\alpha^2 \frac{8\pi}{3}\sum_{i<j} (\bm{s}_i\cdot\bm{s}_j)\,\delta(\bm{r}_{ij}).
\end{equation}
In Eqs.~(\ref{rel:mv})--(\ref{rel:ss}) the indices $i$ and $j$ run over all electrons in the system, $\bm{r}_i$ is the position of the $i$-th electron, $\bm{p}_i=-i\nabla_{\bm{r}_i}$ is the corresponding momentum operator, and $\bm{s}_i=1/2\,\bm{\sigma}_i$ is the spin operator, where $\bm{\sigma}$ is a vector of Pauli matrices. 
The index $I$ runs over all nuclei with charges $Z_I$ located at positions $\bm{r}_I$, while
$\bm{r}_{Ii}=\bm{r}_i-\bm{r}_I$ and $\bm{r}_{ij}=\bm{r}_j-\bm{r}_i$ denote interparticle vectors. Finally, $\delta(\bm{r})$ is the Dirac delta function.

Expectation values of the one-electron relativistic operators, $\hat{H}_\mathrm{mv}$ and $\hat{H}_\mathrm{D1}$, are usually larger in magnitude than the expectation values of the two-electron relativistic operators, $\hat{H}_\mathrm{D2}$ and $\hat{H}_\mathrm{oo}$, but have opposite signs and cancel each other to a large extent \cite{Piszczatowski2008}. 
Therefore, it is advantageous to use their sum -- the Cowan--Griffin correction \cite{Cowan:76} defined as the expectation value of the operator
\begin{equation} \label{rel:CG}
\hat{H}_\mathrm{CG}=\hat{H}_\mathrm{mv}+\hat{H}_\mathrm{D1}.
\end{equation}
In the case of electronic closed-shell singlet states, such as the ground states of argon dimer and argon atom, expectation values of $\hat{H}_\mathrm{D2}$ and $\hat{H}_\mathrm{ssc}$ are related through the formula \cite{Coriani:04}
\begin{equation}
\langle\hat{H}_\mathrm{ssc}\rangle=-2\langle\hat{H}_\mathrm{D2}\rangle.
\end{equation}
As a result, the total relativistic correction for a closed-shell system can be written as
\begin{equation}
E_\mathrm{rel}
=\langle\hat{H}_\mathrm{CG}\rangle
-\langle\hat{H}_\mathrm{D2}\rangle
+\langle\hat{H}_\mathrm{oo}\rangle.
\end{equation}
The relativistic correction to the interaction energy within the supermolecular approach is defined as
\begin{equation} \label{eq:Vrel1}
V_\mathrm{rel} = E_\mathrm{rel}^{\mathrm{Ar}_2} - 2\,E_\mathrm{rel}^\mathrm{Ar},
\end{equation}
where $E_\mathrm{rel}^{\mathrm{Ar}_2}$ and $E_\mathrm{rel}^\mathrm{Ar}$ are the relativistic corrections to the energy of dimer and atom, respectively.
Alternatively, $V_\mathrm{rel}$ may be viewed as a combination of three components calculated separately
\begin{equation} \label{Vrel2}
V_\mathrm{rel}=V_\mathrm{CG}-V_\mathrm{D2}+V_\mathrm{oo},
\end{equation}
where each of the potentials $V_\mathrm{Y}$, $\mathrm{Y}\in\{\mathrm{CG},\mathrm{D2},\mathrm{oo}\}$, is defined analogously as in Eq.~(\ref{eq:Vrel1}), but using the expectation values of the individual operators $\hat{H}_\mathrm{Y}$ instead of $E_\mathrm{rel}$.

In the calculations of the relativistic corrections, we employed fully uncontracted singly-augmented Dunning basis sets aug-cc-pV$X$Z for argon \cite{Dunning1993ar,Dunning1994aug} with cardinal numbers $X$ in the range $X=2\!-\!5$. 
Additionally, a set of midbond functions ($3s3p2d2f1g$) was placed in the middle of the argon-argon bond \cite{slavicek2003,patkowski2010}. 
The basis sets will be denoted $X$Z further in the text. 
We applied the so-called counterpoise scheme \cite{Boys1970} to remove the basis set superposition error, i.e., the energies of both the dimer and the atom in Eq.~(\ref{eq:Vrel1}) were computed in the basis set of the dimer \cite{DCBS95}. 
All calculations were performed using the coupled-cluster method with single, double, and perturbative triple excitations [CCSD(T)] as implemented in the Dalton program \cite{Coriani:04,daltonshort,dalton2018}.
We checked that the effects of core-core and core-valence correlation and the effects of higher excitations are negligible compared to estimated uncertainties of our calculations.
The relativistic corrections were calculated for 37 internuclear distances, including 33 points with $R\ge2$~\AA{} on the same grid as in Ref.~\cite{patkowski2010} and four points ($R=1.2,\,1.4,\,1.6,\,1.8$~\AA) added to improve the accuracy of the potential for small distances.

In order to reduce the basis set incompleteness error in our calculations, and to assess the uncertainty of the \emph{ab initio} data, we employed the Riemann extrapolation scheme introduced in Ref.~\cite{lesiuk2019riemann} and used in our previous study of the helium dimer \cite{czachorowski2020}. 
The two-point Riemann extrapolation formula reads
\begin{equation}\label{eq:riemann}
V_\mathrm{Y}^\infty = V_\mathrm{Y}^X 
+ X^{n_\mathrm{Y}}
\big(V_\mathrm{Y}^X-V_\mathrm{Y}^{X-1}\big)
\big[\zeta(n_\mathrm{Y})-\sum_{i=1}^Xi^{-n_\mathrm{Y}}\big],
\end{equation}
where $\zeta(s) = \sum_{i=1}^\infty i^{-s}$ is the Riemann zeta function, $V_\mathrm{Y}^{X-1}$ and $V_\mathrm{Y}^X$, $\mathrm{Y}\in\{\mathrm{CG},\mathrm{D2}\}$, are values of the corrections calculated using basis sets with cardinal numbers $X-1$ and $X$, respectively, and $V_\mathrm{Y}^\infty$ denotes the corresponding complete basis set limit. 
Further in the text, we denote the extrapolated results from $(X-1)$Z and $X$Z basis sets by the symbol $[(X-1)X]$Z.
The rate of the convergence, characterized by the value of the exponent $n_\mathrm{Y}$ in Eq.~\eqref{eq:riemann}, depends on the correction.
We assumed that the Cowan--Griffin term converges to the basis set limit at the same rate as the BO energy, i.e., we used $n_\mathrm{CG}=4$. 
In the case of the two-electron Darwin term we employed $n_\mathrm{D2}=2$ according to the analytic results of Kutzelnigg \cite{kutzelnigg2008} for the helium atom. 
As discussed further in the text, the orbit-orbit term was not extrapolated.

\begin{figure}
\includegraphics[width=1\columnwidth]{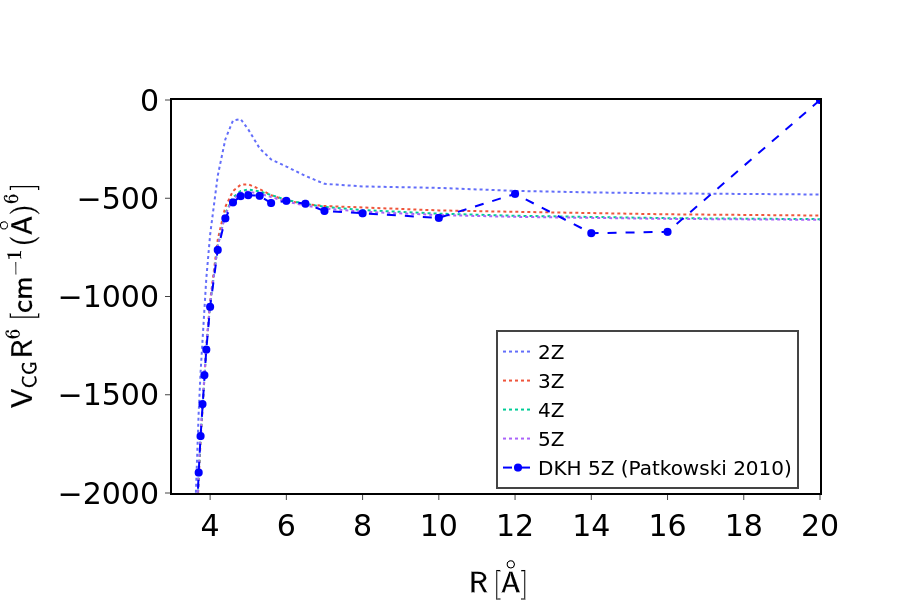}
\caption{The Cowan--Griffin correction, $V_\mathrm{CG}$, to the interaction energy of the argon dimer as a function of the internuclear distance $R$ calculated using $X$Z basis sets. 
The results are multiplied by the sixth power of the internuclear distance to show the convergence to the leading-order term in the large-$R$ asymptotic expansion. 
The DKH results of Ref.~\cite{patkowski2010} are presented for comparison. 
\label{fig:CGR6plot}}
\end{figure}

\begin{figure}
\includegraphics[width=1\columnwidth]{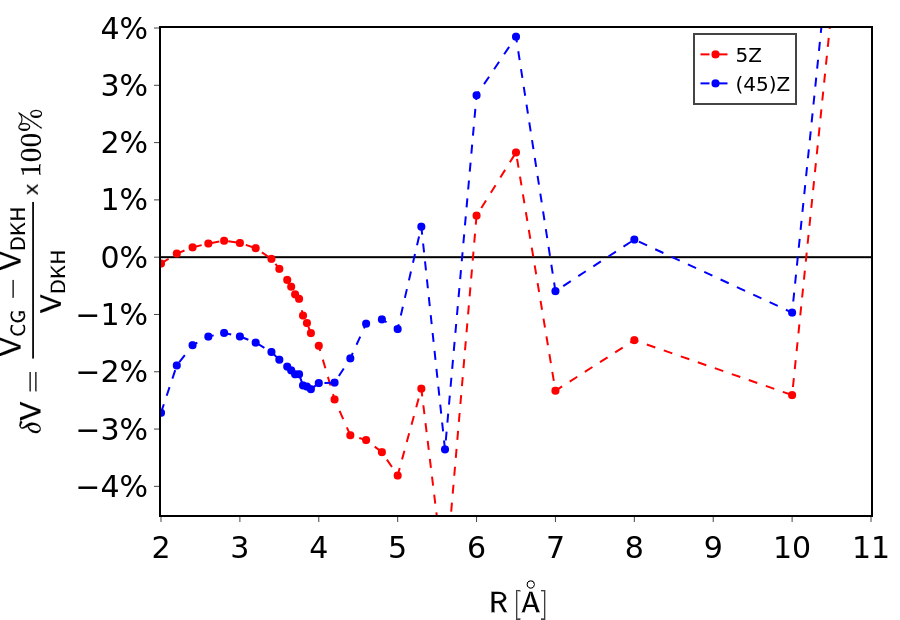}
\caption{Relative percentage difference between the Cowan--Griffin correction calculated in this work and the DKH results from Ref.~\cite{patkowski2010}. 
We compare the data obtained with the 5Z basis set (red points), and results obtained from the (45)Z extrapolation (blue points).
\label{fig:CGplot}}
\end{figure}

In Fig.~\ref{fig:CGR6plot} we present the Cowan--Griffin contribution to the interaction potential calculated in this work. 
The DKH results of Patkowski and Szalewicz \cite{patkowski2010} are given for comparison. 
The apparent divergence of the latter for large $R$ is a numerical artifact originating from insufficient precision of the data from Ref.~\cite{patkowski2010}, where the values were provided with only one or two significant digits. 
Besides this, a good agreement is observed between the DKH method and our Cowan--Griffin results calculated using basis sets with cardinal numbers $X>2$. 
This is clear in Fig.~\ref{fig:CGplot} where we show relative differences between both approaches. 
On average, within the 5Z basis set this difference is about 2.7\% and increases with $R$.
Near the minimum of the potential energy curve ($R=3.75$~\AA) the difference amounts to only 0.7\%. 
As the final values of the Cowan-Griffin contribution we take the (45)Z extrapolated results with the uncertainty estimated as the absolute difference between the 5Z value and the (45)Z extrapolant.
For example, at $R=3.75$~\AA{} the recommended value is $-0.602(8)$~cm$^{-1}$, see Table~\ref{tab:postBO}.

\begin{figure}
\includegraphics[width=1\columnwidth]{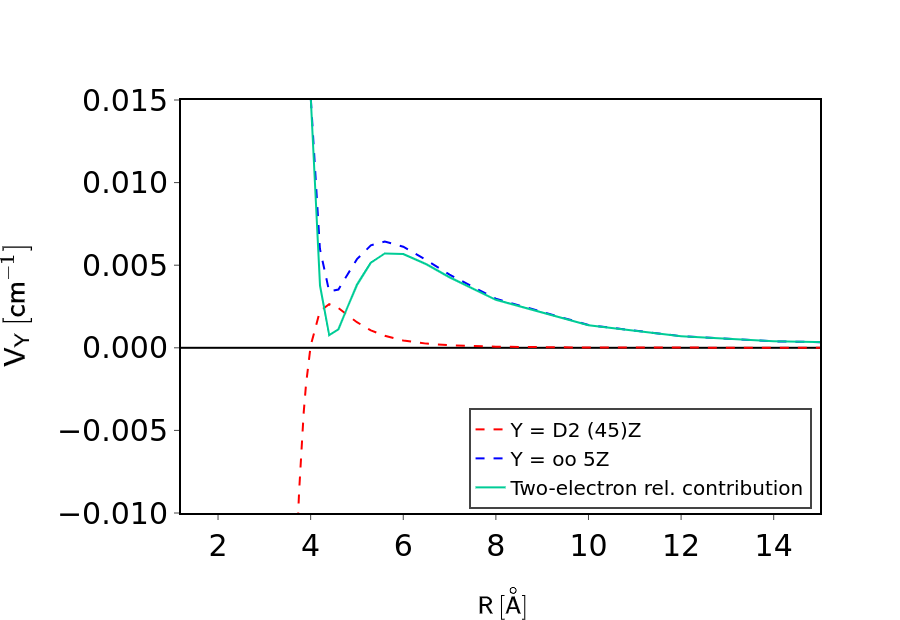}
\caption{Comparison of the two-electron Darwin, $V_\mathrm{D2}$, and orbit-orbit, $V_\mathrm{oo}$, corrections to the interaction energy of the argon dimer and the total two-electron relativistic contribution, $-V_\mathrm{D2}+V_\mathrm{oo}$.
\label{fig:D2plot}}
\end{figure}

In the calculations from Ref.~\cite{patkowski2010}, the two-electron terms in the Breit--Pauli Hamiltonian were neglected and the missing contribution was estimated to be no larger than 15\% of the one-electron correction. 
In this work, we explicitly calculated the two-electron terms for all required internuclear distances. 
Similarly as in the case of the Cowan-Griffin contribution, we consider the extrapolated (45)Z results of the two-electron Darwin correction as converged with the uncertainty estimated as the absolute difference between the 5Z value and the (45)Z extrapolant.
For the orbit-orbit term we encountered non-monotonic convergence of the calculated results with respect to the cardinal number $X$, which makes the reliability of any extrapolation questionable. 
Fortunately, this two-electron correction depends weakly on the basis set size (for $X>2$) and the differences between the results obtained with the 4Z and 5Z basis sets are negligibly small for all considered internuclear distances. 
Therefore, we take the values obtained using the 5Z basis set as the recommended results for the orbit-orbit correction, and the absolute differences between the 4Z and 5Z results are used as the uncertainty estimates. 
In Fig.~\ref{fig:D2plot} we show the final values of the two-electron Darwin and orbit-orbit corrections to the interaction energy as functions of the internuclear distance.
Our value of the total two-electron relativistic correction for the distance $R=3.75$~\AA\ is $0.058(5)$~cm$^{-1}$, see Table~\ref{tab:postBO}, slightly lower than the value of Patkowski and Szalewicz \cite{patkowski2010}. 

For distances up to $R=5$~\AA, the total two-electron relativistic correction amounts to between 5\% and 12\% of the Cowan--Griffin contribution. 
However, with increasing internuclear distance, the effect of slower asymptotic decay of the orbit-orbit term \cite{Meath:66A} starts to be visible, see Table~\ref{tab:postBO}.
At $R=5.3$~\AA, the two-electron terms correspond to about 23\% of the Cowan--Griffin contribution, and finally become larger in magnitude than the latter for $R>7$~\AA. 

The calculated components of the relativistic correction, together with their estimated uncertainties, can be found in Supplemental Material \cite{supp}. The combined uncertainty of the total relativistic correction was obtained as a square root of the sum of squares of the uncertainties of the components. This quantity is also provided in Supplemental Material \cite{supp}.

\begin{table}
\caption{Contributions to the pair potential of argon at the internuclear distance $R=3.75$~\AA, i.e., near the minimum of the potential, and at $R=8$~\AA. 
The energies are in cm$^{-1}$ and the uncertainties at the last digits are given in parentheses.
\label{tab:postBO}}
\begin{ruledtabular}
\begin{tabular}{ld{3.8}d{2.10}}
contribution & 
\multicolumn{2}{c}{value} \\\cline{2-3}
& 
\multicolumn{1}{c}{$R=3.75$~\AA} & 
\multicolumn{1}{c}{$R=8$~\AA} \\
\hline
Born--Oppenheimer\footnote{\label{note:BO}Taken from Ref.~\cite{patkowski2010}.}    
                                  & -98.68(32)   & -1.3238(38)   \\[0.5ex]
Cowan--Griffin                    &  -0.602(8)   & -0.00221(4)   \\
two-electron Darwin $\times(-1)$  &   0.0087(13) & -0.000064(10)   \\
orbit-orbit                       &   0.049(5)   &  0.0030(3)    \\[0.5ex] 
QED                               &   0.0361(3)  &  0.000396(6)  \\
\hline
total  & -99.19(32) & -1.3227(38)
\end{tabular}  
\end{ruledtabular}
\end{table}

\subsection{QED correction \label{subsec:potQED}}

Another important contribution to the interaction potential originates from the QED effects of the order of $\alpha^3$. 
In general, the $\alpha^3$ QED correction to the energy of closed-shell atomic and molecular systems is given by the expression \cite{caswell86,pachucki93,pachucki98}
\begin{equation} \label{anqed}
\begin{split}
E_{\mathrm{QED}(\alpha^3)} 
&= \frac{8\alpha}{3\pi} \left( \frac{19}{30} - 2\ln \alpha - \ln k_0 \right)
   \langle\hat{H}_\mathrm{D1}\rangle \\
&- \frac{\alpha}{\pi} \left( \frac{164}{15} + \frac{14}{3}\ln\alpha \right)
   \langle\hat{H}_\mathrm{D2}\rangle 
 + \langle\hat{H}_\mathrm{AS}\rangle,
\end{split}
\end{equation}
where $\langle\hat{H}_\mathrm{D1}\rangle$ and $\langle\hat{H}_\mathrm{D2}\rangle$ are the expectation values of the relativistic one- and two-electron Darwin operators defined in Eqs.~(\ref{rel:D1}) and (\ref{rel:D2}), respectively, $\langle\hat{H}_\mathrm{AS}\rangle$ is the Araki--Sucher correction \cite{araki57,sucher58}, and $\ln k_0$ is the so-called Bethe logarithm \cite{BeSal,schwartz61}. 
In principle, $\ln k_0$ for the dimer should be computed for each internuclear distance separately. 
However, due to the high computational cost of such calculations, we use a constant value of $\ln k_0$ corresponding to the isolated atoms limit ($R\rightarrow\infty$) and equal to $\ln k_0$ for a single atom (the Bethe logarithm is an intense quantity).
This is expected to be an excellent approximation since $\ln k_0$ depends very weakly on the internuclear distance. 
For example, calculations for the hydrogen molecule in the electronic ground state have shown that $\ln k_0$ changes by less than 1\% when $R$ varies from $1.5$~a.u.\ to infinity \cite{piszczatowski09}.
In this work, we employed the atomic value of the Bethe logarithm for argon,
\begin{equation} \label{anlnk0}
\ln k_0=8.761,
\end{equation} 
taken from Ref.~\cite{lesiuk23bethe}.

The remaining QED effects (of the order of $\alpha^4$ and higher) are usually dominated \cite{Pachucki:06a4,Pachucki:06a4E} by a simple one-loop radiative correction known from the hydrogenic Lamb shift. For atoms and homonuclear dimers it is given by the expression \cite{eides01}
\begin{equation} \label{anqedplus}
E_{\mathrm{QED}(\alpha^4,\text{one-loop})} 
 = 2\alpha^2 Z \left(\frac{427}{96}-2\ln2\right)
   \langle\hat{H}_\mathrm{D1}\rangle,
\end{equation}
where $Z$ is the nuclear charge of an atom. We expect that this correction provides a good estimate of the total $\alpha^4$ QED effect, at least for small and intermediate internuclear distances.

In our treatment of the QED correction to the interaction potential, we included only the effects that are proportional to the relativistic one-electron Darwin correction. 
Combining appropriate terms from Eqs.~(\ref{anqed}) and (\ref{anqedplus}), together with the value of $\ln k_0$ from Eq.~(\ref{anlnk0}), one arrives at the formula
\begin{equation}
V_\mathrm{QED}=0.0165 \, V_\mathrm{D1},
\end{equation}
where the potential $V_\mathrm{D1}$ is constructed from the expectation values of the $\hat{H}_\mathrm{D1}$ operator using an expression similar to Eq.~(\ref{eq:Vrel1}).
Note that due to a large nuclear charge of the argon atom, the contributions to $V_\mathrm{QED}$ arising from Eqs.~(\ref{anqed}) and (\ref{anqedplus}) are of similar magnitude, namely $0.0106\,V_\mathrm{D1}$ and $0.0059\,V_\mathrm{D1}$, respectively, even though they are of a different order in $\alpha$.
The term in Eq.~(\ref{anqed}) proportional to $\langle\hat{H}_\mathrm{D2}\rangle$ can be safely neglected due to small scaling factor, equal to $0.0279$, which makes this term several times smaller than the estimated uncertainty of $V_\mathrm{D2}$, see Table~\ref{tab:postBO}.
The last term appearing in Eq.~(\ref{anqed}), the Araki--Sucher correction, is especially difficult to calculate rigorously \cite{balcerzak2017calculation,lesiuk19a,jaquet20,czachorowski2020}.
Its contribution to the interaction potential for the argon dimer has been evaluated previously only for the minimum of the potential, where it is of the order of $-0.02$~cm$^{-1}$ \cite{balcerzak2017calculation}. 
Therefore, it is safe to assume that for small and intermediate internuclear distances the Araki--Sucher correction is at least an order of magnitude smaller than the dominant source of the uncertainty in the potential (the BO contribution) and is negligible within the accuracy requirements of this work. 
It is known, however, that the Araki--Sucher correction to the potential asymptotically decays as $R^{-3}$ \cite{Meath:66B,Pachucki:05longrange}, and must become important at larger distances. 
This contribution is taken care of by the Casimir--Polder theory and is included in the retardation correction considered by us in Sec.~\ref{subsec:retard}.

To obtain $V_\mathrm{QED}$ we reused data from the calculations of the relativistic corrections described in Sec.~\ref{subsec:potrel}.  
The values computed using the 5Z basis set are treated as the recommended results for the QED correction with the uncertainty estimated as the absolute difference between the 4Z and 5Z values.
The QED correction amounts to 5--20\% of the relativistic Cowan--Griffin correction, with the ratio increasing with $R$, and is larger than the estimated uncertainty of the latter, see Table~\ref{tab:postBO}.
The calculated values of $V_\mathrm{QED}$ can be found in Supplemental Material \cite{supp}.

\section{Fit of the potential \label{sec:fit}}

\subsection{Born--Oppenheimer potential \label{subsec:fitBO}}

As Patkowski and Szalewicz \cite{patkowski2010} fitted only the sum of the BO interaction energy and the DKH correction, we refitted the BO pair potential alone in order to combine it with the data for the post-BO corrections calculated by us. 
The analytic function used in the fitting has the following form
\begin{equation} \label{eq:fitBO}
V_\mathrm{BO}(R)
= \sum_{j=1}^{2}\sum_{i=-1}^{2} a_{ij}R^i\,e^{-\alpha_j R}
- \sum_{k=3}^{8}\frac{C_{2k}^\mathrm{BO}f_{2k+1}(\eta R)}{R^{2k}},
\end{equation}
where $a_{ij}$, $\alpha_i$, and $\eta$ are adjustable parameters, $f_n(x)$ are the Tang--Toennies damping functions \cite{tang1984improved}
\begin{equation}
f_n(x)=1-e^{-x}\sum_{i=0}^n\frac{x^i}{i!},
\end{equation}
and $C_n^\mathrm{BO}$ are the asymptotic constants. 
Using $C_6^\mathrm{BO}$, $C_8^\mathrm{BO}$, and $C_{10}^\mathrm{BO}$ from Ref.~\cite{jiang2015}, the values of higher asymptotic constants, $C_{12}^\mathrm{BO}$, $C_{14}^\mathrm{BO}$, and $C_{16}^\mathrm{BO}$, were computed employing the extrapolation formula of Thakkar \cite{thakkar88}. 
The values of $C_n^\mathrm{BO}$, $n\ge10$, obtained in this way were then fixed in Eq.~(\ref{eq:fitBO}), while $C_6^\mathrm{BO}$ and $C_8^\mathrm{BO}$ were treated as two additional parameters to be fitted.

In the fitting of $V_\mathrm{BO}(R)$, the values of adjustable parameters were constrained by imposing the condition
\begin{equation} \label{eq:fitBOconst}
V_\mathrm{BO}(R) 
= \frac{18^2}{R} 
+ (E^\mathrm{Kr}_\mathrm{BO}-2 E^\mathrm{Ar}_\mathrm{BO}) + \mathcal{O}(R^2),
\end{equation}
that ensures the correct short-range asymptotics of the potential. 
The difference between the ground state energies of krypton (the united-atom limit) and two argon atoms, i.e., $(E^\mathrm{Kr}_\mathrm{BO}-2 E^\mathrm{Ar}_\mathrm{BO})=-1698.4211$~a.u., was calculated at the coupled-cluster singles, doubles, and triples (CCSDT) level of theory \cite{noga1987full} with the aug-cc-pV5Z basis set using the CFOUR program \cite{matthews2020coupled,cfourshort}.
Note that the $C_n^\mathrm{BO}$ coefficients would appear in the three fitting constraints resulting from Eq.~(\ref{eq:fitBOconst}) if the long-range terms $R^{-n}$ in Eq.~(\ref{eq:fitBO}) were damped using the Tang--Toennies functions $f_n$ as it is usually the case.
Since our $C^\mathrm{BO}_6$ and $C^\mathrm{BO}_8$ are not fixed, this would severely complicate the fitting procedure. 
To circumvent this problem, we used a stronger damping of the long-range contributions in Eq.~(\ref{eq:fitBO}), i.e., the $R^{-n}$ terms are multiplied by the functions $f_{n+1}$ instead of $f_n$.

\begin{figure}
\includegraphics[width=1\columnwidth]{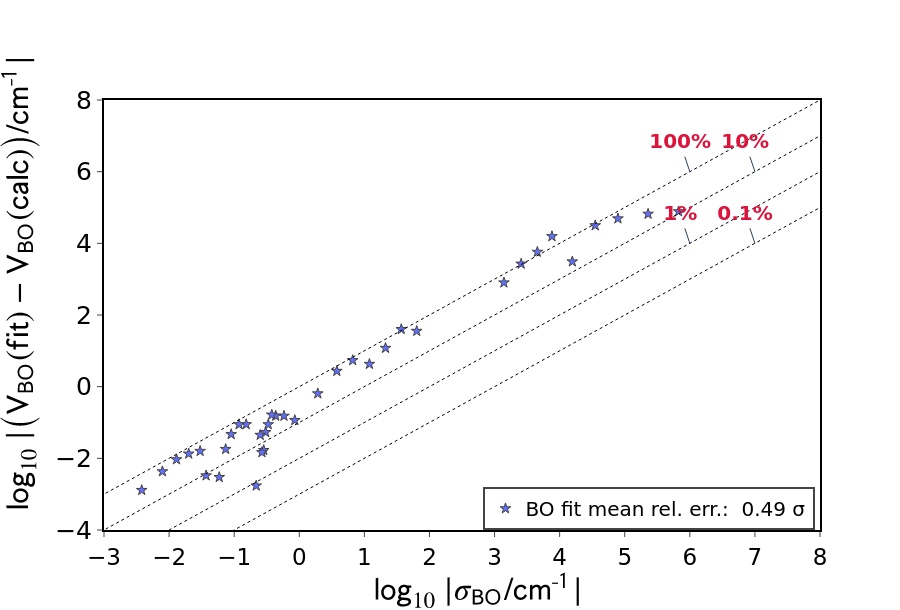}
\caption{Absolute differences between the BO pair potential predicted from the fit, $V_\mathrm{BO}(\mathrm{fit})$, and obtained from \emph{ab initio} calculations, $V_\mathrm{BO}(\mathrm{calc})$, versus estimated theoretical uncertainties $\sigma_\mathrm{BO}$.
The dotted lines and percentage in the legend correspond to relative errors with respect to the estimated uncertainty (1$\sigma_\mathrm{BO}=100\%$).
\label{fig:BOscat}}
\end{figure}

The fitting procedure was performed using the non-linear weighted least squares dogbox algorithm \cite{voglis2004rectangular} with weights equal to the squared inverse of the estimated uncertainties $\sigma_\mathrm{BO}$. 
In Fig.~\ref{fig:BOscat} we show the comparison of absolute errors of the fit with $\sigma_\mathrm{BO}$. 
The mean absolute error of the fit is $0.49\,\sigma_\mathrm{BO}$, i.e., it is about two times smaller than the inherent uncertainty of the \emph{ab initio} data points.
The values of the adjusted and fixed parameters of the $V_\mathrm{BO}(R)$ potential are collected in Table~\ref{tab:fitcoeff}.
Our final value of the $C_6^\mathrm{BO}$ coefficient, $C_6^\mathrm{BO} = 64.2295$~a.u., is similar to the result of Patkowski and Szalewicz, $C_6^\mathrm{BO}=64.2890$~a.u.\ \cite{patkowski2010}, and the estimate of Kumar and Meath, $C_6^\mathrm{BO}=64.30$~a.u.\ \cite{kumar85}. 
Only a slightly worse agreement is obtained for the $C_8^\mathrm{BO}$ coefficient, $C^\mathrm{BO}_8=1503.453$~a.u., in comparison with $C^\mathrm{BO}_8=1514.86$~a.u.\ from Ref.~\cite{patkowski2010} and $C^\mathrm{BO}_8=1621.5$~a.u.\ from Ref.~\cite{jiang2015}.

\begin{table}
\caption{Parameters of analytic fits to the $V_\mathrm{Y}(R)$ potentials, $\mathrm{Y}\in\{\mathrm{BO},\mathrm{rel},\mathrm{QED}\}$, defined in Eqs.~(\ref{eq:fitBO}), (\ref{eq:fitrel}), and (\ref{eq:fitQED}). 
The symbol $A(p)$ means $A\times10^p$.
\label{tab:fitcoeff}}
\begin{ruledtabular}
\begin{tabular}{cd{2.12}d{2.13}d{2.12}}
& 
\multicolumn{1}{c}{$\mathrm{Y}=\mathrm{BO}$} & 
\multicolumn{1}{c}{$\mathrm{Y}=\mathrm{rel}$} &
\multicolumn{1}{c}{$\mathrm{Y}=\mathrm{QED}$} \\ 
\hline
$\alpha_1$ & 6.92639712 & 1.66833595 & 1.64936036 \\
$\alpha_2$ & 2.34781022 & 3.07137482 & 2.29848003 \\
$\eta$     & 3.65946625 & 1.05249333 & 3.91638946 \\[1ex]
$a_{-11}$ & 2.19578828(4) & \multicolumn{1}{c}{-} & \multicolumn{1}{c}{-} \\
$a_{01}$  & 6.77672149(4) &-3.55673715     &-1.2621862(-1) \\
$a_{11}$  & 8.52129476(4) & 7.59367220(-1) & 1.176436(-2)  \\
$a_{21}$  & 9.53252754(4) &-4.19874548(-2) & \multicolumn{1}{c}{-} \\
$a_{-12}$ &-2.16338828(4) & \multicolumn{1}{c}{-} & \multicolumn{1}{c}{-} \\
$a_{02}$  & 3.18311293(4) &-2.83560129(1)  & \multicolumn{1}{c}{-} \\
$a_{12}$  &-8.18603449(3) & 1.64892143(1)  & \multicolumn{1}{c}{-} \\
$a_{22}$  & 8.37564742(2) & 7.19542232     & 6.921721(-2)  \\[1ex]
$C_4^Y$   & \multicolumn{1}{c}{-} & -9.43594504(-4) & \multicolumn{1}{c}{-} \\
$C_6^Y$    &6.42295222(1)  & 1.76276088(-1)  & -1.949525(-2) \\
$C_8^Y$    &1.50345262(3)  & 3.99787449(-1)  & -4.8877359(-1) \\
$C_{10}^Y$ &4.9033(4)      & \multicolumn{1}{c}{-} & \multicolumn{1}{c}{-} \\
$C_{12}^Y$ &1.828012(6)    & \multicolumn{1}{c}{-} & \multicolumn{1}{c}{-} \\
$C_{14}^Y$ &8.1913453(7)   & \multicolumn{1}{c}{-} & \multicolumn{1}{c}{-} \\
$C_{16}^Y$ &4.294913366(9) & \multicolumn{1}{c}{-} & \multicolumn{1}{c}{-} \\
\end{tabular}
\end{ruledtabular}
\end{table}

\subsection{Relativistic potential \label{subsec:fitrel}}

The relativistic correction to the interaction energy was fitted using the analytic formula
\begin{equation} \label{eq:fitrel}
V_\mathrm{rel}(R)
=\sum_{j=1}^{2}\sum_{i=0}^{2} a_{ij}R^i\,e^{-\alpha_j R}
- \sum_{k=2}^{4}\frac{C^\mathrm{rel}_{2k}f_{2k}(\eta R)}{R^{2k}},
\end{equation}
where $a_{ij}$, $\alpha_i$, and $\eta$ are adjustable parameters.
As no values of the relativistic asymptotic constants for the argon dimer have been published to date, the $C_n^\mathrm{rel}$ coefficients were also treated as fitting parameters.
We restricted the asymptotic expansion to only three leading terms, with coefficients $C^\mathrm{rel}_4$, $C^\mathrm{rel}_6$, and $C^\mathrm{rel}_8$, to reduce the possibility of over-fitting. 

In the fitting of $V_\mathrm{rel}(R)$, we imposed only one constraint on the values of adjustable parameters,
\begin{equation}
V_\mathrm{rel}(0) = (E^\mathrm{Kr}_\mathrm{rel}-2E^\mathrm{Ar}_\mathrm{rel}),
\end{equation}
which arises from the united-atom limit.
The value of $(E^\mathrm{Kr}_\mathrm{rel}-2E^\mathrm{Ar}_\mathrm{rel}) = -31.91275$~a.u.\ was calculated at the CCSDT level of theory with the fully uncontracted aug-cc-pV5Z basis set using the CFOUR program.
This value does not include the orbit-orbit correction which is not available in the CFOUR code.
The fitting procedure was performed as described is Sec.~\ref{subsec:fitBO}.
The mean absolute error of our final fit is $0.05\,\sigma_\mathrm{rel}$, see Fig.~\ref{fig:relqedscat}. 
The determined parameters of the $V_\mathrm{rel}(R)$ potential are reported in Table~\ref{tab:fitcoeff}.

To allow for a direct comparison of our results with the previous work of Patkowski and Szalewicz \cite{patkowski2010}, we additionally constructed an analytic fit of the Cowan--Griffin correction alone, $V_\mathrm{CG}(R)$. 
The functional form of $V_\mathrm{CG}(R)$ is similar to that in Eq.~(\ref{eq:fitrel}) but the $R^{-4}$ term is absent as it arises from the orbit-orbit interaction that is missing in the Cowan--Griffin theory.

\subsection{QED potential \label{subsec:fitQED}}

\begin{figure}
\includegraphics[width=1\columnwidth]{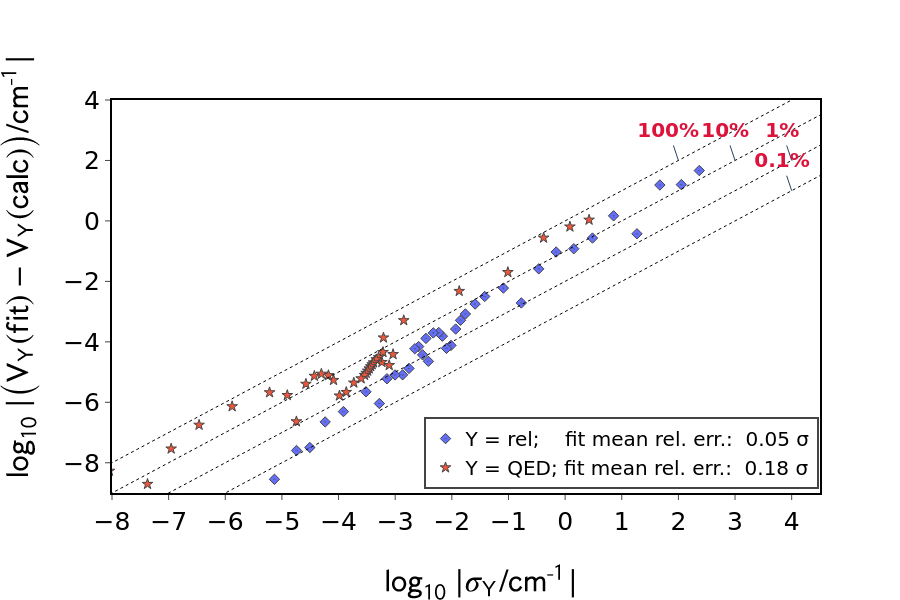}
\caption{Absolute differences between the corrections to the interaction potential $V_\mathrm{Y}$, $\mathrm{Y}\in\{\mathrm{rel},\mathrm{QED}\}$, predicted from the fit, $V_\mathrm{Y}(\mathrm{fit})$, and obtained from \emph{ab initio} calculations, $V_\mathrm{Y}(\mathrm{calc})$, versus estimated theoretical uncertainties $\sigma_\mathrm{Y}$. 
The dotted lines and percentage in the legend correspond to relative errors with respect to the estimated uncertainty (1$\sigma_\mathrm{Y}=100\%$).
\label{fig:relqedscat}}
\end{figure}

The QED correction to the interaction energy was fitted using a model function
\begin{equation} \label{eq:fitQED}
\begin{split}
V_\mathrm{QED}(R) 
& = (a_{01}+a_{11} R )e^{-\alpha_1 R} + a_{22} R^2e^{-\alpha_2 R}\\
& - \sum_{k=3}^{4}\frac{C^\mathrm{QED}_{2k}f_{2k}(\eta R)}{R^{2k}},
\end{split}
\end{equation}
where $a_{01}$, $a_{11}$, $a_{22}$, $\alpha_1$, $\alpha_2$, $\eta$, $C_6^\mathrm{QED}$, and $C_8^\mathrm{QED}$ are all adjustable parameters.
No constraints on their values were imposed during the fitting procedure.
The mean absolute error of the final fit is $0.18\,\sigma_\mathrm{QED}$, see Fig.~\ref{fig:relqedscat}. 
The determined parameters of the $V_\mathrm{QED}(R)$ potential are shown in Table~\ref{tab:fitcoeff}.

\subsection{Long-range retardation of the potential \label{subsec:retard}}

In the region where the internuclear distances become very large, the retardation of the electromagnetic radiation has to be taken into account. 
According to the theory of Casimir and Polder \cite{casimir1948}, the retarded pair potential for the argom dimer for large $R$ is given by
\begin{equation} \label{eq:cp}
V_\mathrm{CP}(R) = -\frac{1}{\pi R^6}
\int_0^\infty [\alpha_1(i\omega)]^2 e^{-2\alpha\omega R} P(\alpha\omega R)~d\omega, 
\end{equation}
where $\alpha_1(i\omega)$ is the dynamic dipole polarizability of the argon atom at an imaginary frequency $i\omega$ and $P(x)=x^4+2x^3+5x^2+6x+3$. 
Performing the expansion of $V_\mathrm{CP}(R)$ for a fixed $R$ in powers of $\alpha$ through the $\alpha^3$ terms one obtains \cite{Meath:66B,Pachucki:05longrange}
\begin{equation} \label{eq:cpshort}
V_\mathrm{CP}(R)=-\frac{C_6}{R^6}-\frac{C_4}{R^4}-\frac{C_3}{R^3}+\mathcal{O}(\alpha^4),
\end{equation}
where the consecutive terms on the right-hand-side of Eq.~(\ref{eq:cpshort}) represent the leading terms in the asymptotic expansion (in $R^{-1}$) of the BO potential, the orbit-orbit contribution to the $\alpha^2$ relativistic correction, and the Araki--Sucher contribution to the $\alpha^3$ QED correction, respectively.
The appropriate powers of $\alpha$ are included in the definitions of the asymptotic constants $C_6$, $C_4$, and $C_3$, which can be found in Refs.~\cite{Meath:66B,Pachucki:05longrange}.
In particular, the formula for the $C_3$ constant for the argon dimer reads
\begin{equation} \label{eq:cpshortC3}
C_3=\frac{7\alpha^3}{6\pi}18^2,
\end{equation}
where the factor $18^2$ comes from the nuclear charge.
At large $R$, the Casimir--Polder potential $V_\mathrm{CP}(R)$ behaves as \cite{casimir1948,Meath:66B}
\begin{equation} \label{eq:cplong}
V_\mathrm{CP}(R)=-\frac{K_7}{R^7}+\mathcal{O}(R^{-9}),
\end{equation}
where
\begin{equation} \label{eq:cplongK7}
K_7=\frac{23}{4\pi\alpha}[\alpha_1(0)]^2.
\end{equation}
As a result, the dominant effect of the retardation is to switch the usual London’s $R^{-6}$ long-range decay of the interaction energy to the $R^{-7}$ form.

Following the approach developed in the study of the helium dimer and presented in Refs.~\cite{przybytek2010,cencek2012,przybytek2012}, we introduce the effects of retardation to the potential through an additive correction $\delta V_\mathrm{ret}(R)$.
Our total interaction potential of the argon dimer, $V(R)$, is thus defined as a sum
\begin{equation} \label{eq:Vtot}
V(R)=V_\mathrm{BO}(R)+V_\mathrm{rel}(R)+V_\mathrm{QED}(R)+\delta V_\mathrm{ret}(R),
\end{equation}
where
\begin{equation} \label{eq:Vtotret}
\delta V_\mathrm{ret}(R)=V_\mathrm{CP}(R)+\frac{C_6}{R^6}+\frac{C_4}{R^4}.
\end{equation}
This particular form of the retardation correction is a consequence of the fact that we explicitly include both the BO interaction energy and the orbit-orbit correction in our \emph{ab initio} data, and the leading terms in the asymptotic expansion of these interactions are present in $V_\mathrm{BO}(R)$ and $V_\mathrm{rel}(R)$, respectively. 
Therefore, $-C_6/R^6-C_4/R^4$ must be subtracted from $V_\mathrm{CP}(R)$ to prevent double counting, see Eq.~(\ref{eq:cpshort}).
Moreover, although our $V_\mathrm{QED}(R)$ potential does not include the Araki--Sucher correction, its leading asymptotic term, $-C_3/R^3$, is taken into account for small and intermediate internuclear distances when the retardation correction is applied.
It must be pointed out that the addition of $\delta V_\mathrm{ret}(R)$ does not entirely eliminate all unphysical long-range terms from our potential. 
The very small $R^{-6}$ terms in $V_\mathrm{rel}(R)$ and $V_\mathrm{QED}(R)$ can be eliminated at large $R$ using the relativistic extension of the Casimir--Polder theory proposed by Pachucki \cite{Pachucki:05longrange}. 
However, this is not expected to impact the potential $V(R)$ in the physically relevant range of distances, and was not attempted.

\begin{table}
\caption{Parameters of the rational fit to the retardation factor $g(R)$ defined in Eq.~(\ref{eq:ratg}). 
The symbol $A(p)$ means $A\times10^p$.
\label{tab:fitratg}}
\begin{ruledtabular}
\begin{tabular}{cd{1.21}d{1.21}}
$i$ & \multicolumn{1}{c}{$A_i$} & \multicolumn{1}{c}{$B_i$} \\ 
\hline
1& 3.064827731712776(-1)  & 3.064827731712776(-1)  \\
2& 4.843969950480238(-3)  & 4.858660928580025(-3)  \\
3& 7.793815409352866(-6)  & 1.156839432134599(-5)  \\
4& 1.031221966985517(-14) & 1.670845939899731(-8)  \\
5& 4.204140311858030(-12) & 2.151070686307512(-17) \\
6& \multicolumn{1}{c}{-}  & 8.769598859882065(-15) \\
\end{tabular}
\end{ruledtabular}
\end{table}

The Casimir--Polder potential can be conveniently represented through the retardation factor $g(R)$ defined as \cite{przybytek2010,cencek2012}
\begin{equation} \label{eq:cpg}
V_\mathrm{CP}(R)=-g(R)\frac{C_6}{R^6}.
\end{equation}
We calculated the values of $g(R)$ according to Eq.~(\ref{eq:cp}) for $100$ internuclear distances within the interval $0 \le R \le 250$~a.u.\ using dipole polarizabilities $\alpha_1(i\omega)$ from Ref.~\cite{jiang2015} and $C_6=64.30$~a.u.\ derived from the same data.
The retardation factor was then fitted using a rational function of the form
\begin{equation}
\label{eq:ratg}
g(R) = \frac{1 + \sum_{i=1}^5 A_i R^i}{1 + \sum_{i=1}^6 B_i R^i},
\end{equation}
where $B_i$ are the only adjustable parameters. 
The parameters $A_i$ are constrained in order to assure the correct behavior of $V_\mathrm{CP}(R)$ at both $R\to0$ and $R\to\infty$ limits, defined by Eqs.~(\ref{eq:cpshort}) and (\ref{eq:cplong}), respectively.
The resulting constraints have the following form:
\begin{gather}
\label{eq:A1} A_1 = B_1,\\
\label{eq:A2} A_2 = B_2 + \frac{C_4}{C_6},\\
\label{eq:A3} A_3 = B_3 + B_1\frac{C_4}{C_6} + \frac{C_3}{C_6},\\
\label{eq:A4} A_4 = B_5\frac{K_7}{C_6},\\
\label{eq:A5} A_5 = B_6\frac{K_7}{C_6}.
\end{gather}
In the calculations, summarized in Table~\ref{tab:fitratg}, we used the exact value of $C_3$ defined in Eq.~(\ref{eq:cpshortC3}) and the value of $K_7$, Eq.~(\ref{eq:cplongK7}), computed using the static polarizability of the argon atom, $\alpha_1(0) = 11.08$~a.u., from Ref.~\cite{jiang2015}. 
The constants $C_6$ and $C_4$ were approximated by the coefficients $C_6^\mathrm{BO}$ and $C_4^\mathrm{rel}$ obtained from fitting of the potentials $V_\mathrm{BO}(R)$ and $V_\mathrm{rel}(R)$, respectively, described in the previous sections.
Exactly the same values of $C_6$ and $C_4$ must then be used in Eqs.~(\ref{eq:Vtotret}) and (\ref{eq:cpg}) to construct the retardation correction, $\delta V_\mathrm{ret}(R)$, which is added to the sum of unretarded potentials according to Eq.~(\ref{eq:Vtot}).

In practical implementation, it is advantageous to incorporate the effect of adding $\delta V_\mathrm{ret}(R)$ directly into the functional form of the remaining components of $V(R)$, rather than calculating $\delta V_\mathrm{ret}(R)$ separately. In this way, a possible loss of numerical accuracy for large internuclear distances is prevented when two contributions of a similar magnitude but of opposite signs cancel out to a large extent. This cancellation occurs between the unretarded potential and the retardation correction which both vanish as $R^{-4}$, but their sum forming the retarded potential $V(R)$ vanishes at a significantly faster rate with $R$. To this end, we first introduce a complementary damping function $\tilde{f}_n(x)$ defined as
\begin{equation}
\tilde{f}_n(x)=f_n(x)-1=-e^{-x}\sum_{i=0}^n\frac{x^i}{i!},
\end{equation}
which vanishes when $x$ goes to infinity. The potentials are then modified as follows:
\begin{itemize}
    \item the term $-f_7(\eta R)\,C_6^\mathrm{BO}/R^6$ in $V_\mathrm{BO}(R)$ from Eq.~(\ref{eq:fitBO}) is replaced by $-(g(R)+\tilde{f}_7(\eta R))\,C_6^\mathrm{BO}/R^6$,
    \item the term $-f_4(\eta R)\,C_4^\mathrm{rel}/R^4$ in $V_\mathrm{rel}(R)$ from Eq.~(\ref{eq:fitrel}) is replaced by $-\tilde{f}_4(\eta R)\,C_4^\mathrm{rel}/R^4$.
\end{itemize}
Note that these modifications necessitate no changes to the fitting parameters in the potentials $V_\mathrm{BO}(R)$ and $V_\mathrm{rel}(R)$.

\subsection{Fit of the local uncertainties \label{subsec:fitsigma}}

In order to estimate the uncertainties of physical properties of the argon gas calculated using our potential, we generated fits for functions $\sigma_\mathrm{Y}(R)$, $\mathrm{Y}\in\{\mathrm{BO},\mathrm{rel}\}$, representing the uncertainties of the potentials $V_\mathrm{Y}(R)$ due to the uncertainties of the \emph{ab initio} calculations. 
The estimated uncertainties of the QED correction can be safely neglected as they are much smaller than the uncertainties of both the BO interaction energy and the relativistic correction.
The exact value of the potential $V_\mathrm{Y}(R)$ is expected to be contained within the range $V_\mathrm{Y}(R) \pm \sigma_\mathrm{Y}(R)$ with probability corresponding to $k=2$ coverage. 
Note that the functions $\sigma_\mathrm{Y}(R)$ are not intended to precisely fit the estimated \emph{ab initio} uncertainties but rather to follow general trends in their behavior as functions of the internuclear distance, additionally providing upper bounds for the calculated data. 
The uncertainty function $\sigma(R)$ of the total interaction potential $V(R)$ is obtained by summing squares of the uncertainties $\sigma_\mathrm{BO}(R)$ and $\sigma_\mathrm{rel}(R)$ and taking the square root.

The analytic form of the functions $\sigma_\mathrm{Y}(R)$ used in the fitting reads
\begin{equation}
\label{eq:fitsigma}
\sigma_\mathrm{Y}(R) = s_0\,e^{-\alpha_0 R} + \sum_{i=1}^{3} s_i\,e^{-\alpha_i R^2},
\end{equation}
where $s_0$, $s_i$, and $\alpha_i$ are adjustable parameters, and $\alpha_0$ is fixed [$\alpha_0=1.9$ for $\sigma_\mathrm{BO}(R)$ and $\alpha_0=1.05$ for $\sigma_\mathrm{rel}(R)$] to provide a baseline for the long-range decay of the uncertainty. 
The fitting procedure was performed using the standard least-squares method applied to a reduced set of data points obtained by discarding points where the value of the calculated uncertainty was significantly smaller than for two closest neighboring points. 
The average ratio of the uncertainty predicted from $\sigma_\mathrm{Y}(R)$ to the actual uncertainty of the \emph{ab initio} data is 1.08 for $\sigma_\mathrm{BO}(R)$ and 1.17 for $\sigma_\mathrm{rel}(R)$.
The median of this ratio is 1.16 for both fits. 
The determined parameters of the $\sigma_\mathrm{Y}(R)$ functions are shown in Table~\ref{tab:fitsigma}.

A numerical implementation of all functions discussed in this section, namely $V_\mathrm{Y}(R)$, $\mathrm{Y}\in\{\mathrm{BO},\mathrm{rel},\mathrm{QED}$\}, and $\sigma_\mathrm{Y}(R)$, $\mathrm{Y}\in\{\mathrm{BO},\mathrm{rel}\}$, can be found in Supplemental Material~\cite{supp} as a Fortran 2008 program.

\begin{table}
\caption{Parameters of the uncertainty functions $\sigma_\mathrm{Y}(R)$, $\mathrm{Y}\in\{\mathrm{BO},\mathrm{rel}\}$, defined in Eq.~(\ref{eq:fitsigma}). 
The symbol $A(p)$ means $A\times10^p$.
\label{tab:fitsigma}}
\begin{ruledtabular}
\begin{tabular}{cd{1.13}d{1.14}}
& 
\multicolumn{1}{c}{$\mathrm{Y}=\mathrm{BO}$} & 
\multicolumn{1}{c}{$\mathrm{Y}=\mathrm{rel}$} \\ 
\hline
$\alpha_0$ & 1.9            & 1.05 \\
$\alpha_1$ & 6.81313900(-1) & 1.16451516(-2) \\
$\alpha_2$ & 5.76343421(-3) & 2.58221232(-1) \\
$\alpha_3$ & 2.56907367(-2) & 2.00658289(-3) \\[1ex]
$s_0$ & 6.29130076(-1) & 6.76919448(-5) \\
$s_1$ & 9.85423065(-1) & 1.95220703(-8) \\
$s_2$ & 2.56799977(-8) & 1.89517999(-3) \\
$s_3$ & 2.93856722(-6) & 6.75730842(-10) \\
\end{tabular}
\end{ruledtabular}
\end{table}

\section{Computational methods}

\subsection{Nuclear dynamics calculations \label{subsec:methods_dynamics}}

To calculate the spectroscopic and thermophysical properties of the argon dimer, one has to solve the radial nuclear Schr\"odinger equation 
\begin{equation}\label{eq:schrod}
\left[
  \frac{d^2}{dR^2} - \frac{l(l+1)}{R^2} 
+ \frac{2\mu_a}{\hbar^2} \big(E-U(R)\big)
\right]\chi_l(R) = 0,
\end{equation}
where $l$ is the rotational quantum number, $U(R)$ is the electronic interaction potential, and $\mu_a$ is the reduced mass of the system calculated using atomic masses. 
In the case of the $^{40}\!\mathrm{Ar}_2$ molecule, the latter quantity is equal to $\mu_a=36\,423.5\,m_e$ \cite{mass21I,mass21II}. 
For $E<0$, the solutions $\chi_l(R)$ that fulfill appropriate boundary conditions are the radial wave functions of bound rovibrational states of the molecule and the corresponding values of $E$ are the binding energies denoted $E_{v,l}$, where $v$ is the vibrational quantum number.
The binding energies may be found, for example, through the bisection of the energy variable.
For arbitrary $E>0$, the function $\chi_l(R)$ represents the scattering state and $E$ is the relative collision energy of the system.
The wave function of any scattering state asymptotically approaches the free-particle solution \cite{Joachain75book}
\begin{equation}\label{eq:phase_shift_wf}
\chi_l(R) \approx R
\big( j_l\big(\kappa R)-y_l(\kappa R)\;\mathrm{tan}\,\delta_l(E) \big),
\end{equation}
where $\kappa = \sqrt{2\mu_{a}E}/\hbar$, $j_l(x)$ and $y_l(x)$ are the spherical Bessel functions of the first and second kind, respectively, and $\delta_l(E)$ is the phase shift. 

Several numerical approaches are available to solve second-order differential equations such as Eq.~(\ref{eq:schrod}).
One of the most common techniques is the Numerov integration method \cite{Numerov24,Numerov27,blatt1967}, in particular its renormalized form \cite{johnson1977} which is employed in this work. 
The wave function obtained by the traditional Numerov method can grow exponentially in the classically-forbidden regions which is numerically problematic. 
However, in the renormalized Numerov method one propagates the ratio of the values of the wave function at two consecutive grid points. 
This ratio is, in practice, bounded and hence the problem of the exponential growth is eliminated. 
However, this comes at a price of losing the correct normalization of the wave function which must be restored afterwards. 
Another advantage of the renormalized Numerov method is the direct access to the quantity $\chi_l'(R)/\chi_l(R)$ required in the calculation of phase shifts discussed in Sec.~\ref{subsec:methods_shift}.

\subsection{Second pressure and acoustic virial coefficients \label{subsec:methods_virial}}

The second virial coefficient $B(T)$ can be expressed as a sum of three distinct parts \cite{Beth1937,hirschfelder1964molecular}
\begin{equation}\label{eq:virial}
B(T)= B_{\mathrm{ideal}}(T)+B_{\mathrm{bound}}(T)+B_{\mathrm{thermal}}(T),
\end{equation}
where $B_{\mathrm{ideal}}(T)$ is the ideal gas contribution, $B_{\mathrm{bound}}(T)$ is the contribution from the bound rovibrational states of the dimer, and $B_{\mathrm{thermal}}(T)$ is the atom-atom scattering contribution. 
For the gas of bosonic $^{40}\!$Ar atoms with zero nuclear spin, these contributions are defined as 
\begin{equation}\label{eq:ideal}
B_{\mathrm{ideal}}(T) 
= -\frac{\Lambda(T)^3}{16},
\end{equation}
\begin{equation}\label{eq:bound}
B_{\mathrm{bound}}(T) 
= -\Lambda(T)^3 \sum_{v,l}^{l\; \mathrm{even}}(2l+1)\big(e^{-E_{v,l}/k_\mathrm{B}T}-1\big),
\end{equation}
\begin{equation}\label{eq:thermal}
B_{\mathrm{thermal}}(T) 
= -\frac{\Lambda(T)^3}{\pi k_\mathrm{B}T}\int_0^\infty e^{-E/k_\mathrm{B}T}S(E)~dE,
\end{equation}
where
\begin{equation}\label{eq:LambdaT}
\Lambda(T) = \frac{h}{\sqrt{2\pi \mu_a k_\mathrm{B}T}},
\end{equation}
and
\begin{equation}\label{eq:S_E}
S(E) = \sum_{l}^{l\; \mathrm{even}} (2l+1) \,\delta_l(E).
\end{equation}

The second acoustic virial coefficient $\beta(T)$ is defined using $B(T)$ and its first and second derivatives with respect to the temperature \cite{hirschfelder1964molecular}
\begin{equation}\label{eq:acoustic}
\beta(T) 
= 2B(T) 
+ 2(\gamma_0-1)TB'(T)
+ \frac{(\gamma_0-1)^2}{\gamma_0}T^2B''(T),
\end{equation}
where $\gamma_0$ is the heat capacity ratio ($\gamma_0 = 5/3$ for a monatomic gas).

\subsection{Transport properties \label{subsec:methods_transport}}

The transport coefficients can be obtained using the Chapman--Enskog method applied to the Boltzmann equation \cite{chapman1970,ferziger1973} with the help of the Sonine polynomial expansions as described in detail in Refs.~\cite{loyalka2007,tipton2009a,tipton2009b}. 
In this work, we consider two transport coefficients, namely the thermal conductivity, $\lambda(T)$, and the viscosity, $\eta(T)$. 
In the framework of Chapman and Enskog, the formula for the thermal conductivity is given by
\begin{equation}\label{eq:cond}
\lambda(T) = 
\frac{75}{64}
\frac{k_\mathrm{B}h}{2\mu_a\Lambda(T)}
\frac{f_\lambda^{(m)}}{\Omega^{(2,2)}(T)},
\end{equation}
and for the viscosity by
\begin{equation}\label{eq:visc}
\eta(T) = 
\frac{5}{16}
\frac{h}{\Lambda(T)}
\frac{f_\eta^{(m)}}{\Omega^{(2,2)}(T)},
\end{equation}
where $\Lambda(T)$ is defined in Eq.~(\ref{eq:LambdaT}) and $\Omega^{(2,2)}(T)$ is the collision integral of the order $(2,2)$. 
In general, the integrals $\Omega^{(n,s)}(T)$ are defined as
\begin{equation}\label{eq:omega}
\begin{split}
\Omega^{(n,s)}(T) 
&= \frac{(n+1)}{n(s+1)!(k_\mathrm{B}T)^{s+2}} \\
&\times \int_0^\infty Q^{(n)}(E)\,e^{-E/k_\mathrm{B}T}E^{s+1}~dE,
\end{split}
\end{equation}
and depend on the quantum collision cross-sections
\begin{equation}\label{eq:Q_n}
\begin{split}
Q^{(n)}(E) 
&= \frac{8\pi}{\kappa^2} \sum_{l}^{l\;\mathrm{even}} \sum_{j=0}^{\lfloor(n-1)/2\rfloor} \\
&\times C_{lj}^{(n)} \, \mathrm{sin}^2(\delta_l(E) - \delta_{l+n-2j}(E)),
\end{split}
\end{equation}
where $\lfloor x \rfloor$ denotes the floor function and the coefficients $C_{lj}^{(n)}$ can be found in Refs.~\cite{meeks1994,Sharipov2017}.
$f_\lambda^{(m)}$ and $f_\eta^{(m)}$ appearing in Eqs.~(\ref{eq:cond}) and (\ref{eq:visc}) are correction factors representing $m$-th order approximations of the kinetic theory \cite{hirschfelder1964molecular}. 
In the first-order approximation we have $f_\lambda^{(1)}=f_\eta^{(1)}=1$, and higher-order approximations are close to this value. 
In general, $f_\lambda^{(m)}$ and $f_\eta^{(m)}$ depend on the collision integrals $\Omega^{(n,s)}(T)$; explicit expressions can be found in Ref.~\cite{viehland1995}. 
We employ the fifth-order approximation in this work. 

\subsection{Phase shifts calculations \label{subsec:methods_shift}}

Calculations of the thermophysical properties require the knowledge of two quantities: $S(E)$ defined in Eq.~(\ref{eq:S_E}) and $Q^{(n)}(E)$ defined in Eq.~(\ref{eq:Q_n}). Both are expressed in terms of phase shifts $\delta_l(E)$ which are calculated according to the procedure described in this section.

Using the asymptotic form of the scattering wave function, Eq.~(\ref{eq:phase_shift_wf}), phase shifts can be obtained as a limit \cite{Joachain75book}
\begin{equation}\label{eq:phase_shift-0}
\delta_l(E) = \lim_{R\to\infty} \delta_l(E,R),
\end{equation}
where 
\begin{equation}\label{eq:phase_shift}
\tan\delta_l(E,R) = 
\frac
{\kappa j_l'(\kappa R) - \gamma_l(R)j_l(\kappa R)}
{\kappa y_l'(\kappa R) - \gamma_l(R)y_l(\kappa R)},
\end{equation}
and $\gamma_l(R)$ is calculated from the solution to Eq.~(\ref{eq:schrod}) as
\begin{equation}
\gamma_l(R) = \frac{\chi_l'(R)}{\chi_l(R)}-\frac{1}{R}.
\end{equation}
Due to the periodic nature of the tangent function, from Eq.~(\ref{eq:phase_shift}) we have direct access only to principal values of the phase shifts that are always within the interval $[-\pi/2 ,\pi/2]$. This is sufficient for the calculation of $Q^{(n)}(E)$, but absolute phase shifts are needed to construct $S(E)$. The absolute values may, however, differ from the principal values by an integer multiple of the period length $k\pi$, where $k$ is unknown. In order to find the appropriate value of $k$, we assume that $\delta_l(E,R)$ is a continuous function of $R$ with the initial value $\delta_l(E,0) = 0$, i.e., we assume that $k=0$ for $R=0$. Starting from this condition, $\tan\delta_l(E,R)$ is calculated increasing the internuclear distance in small steps, $\Delta R$. 
When $\tan\delta_l(E,R)$ experiences a jump from positive to negative values at two consecutive $R$, indicating that the period boundary has been crossed, the current value of $k$ is increased by one. 
The opposite procedure is applied, i.e., $k$ is decreased by one, when $\tan\delta_l(E,R)$ jumps from negative to positive values. Alternative approaches to calculating the absolute phase shifts have been given in Refs.~\cite{wei2006,piel2018,palov2021vpa}. 

Initially, we performed calculations using the integration step equal to $\Delta R = 0.5\,E^{-1/3}\times10^{-6}$~a.u. 
This step size is sufficient in the case of $s$-wave phase shifts for small energies $E<10^{-8}$~a.u. 
However, it becomes too large to identify in a reliable way all sign changes of $\delta_l(E,R)$ for states with $l>0$. 
In this case the integration step should be at least two orders of magnitude smaller than for $l=0$. 
This makes the approach described above impractical for routine calculations. 
For this reason, the actual calculations were carried through the method described in Ref.~\cite{wei2006} with the step size given above. 
For energies $E>10^{-3}$~a.u.\ the step size was increased by an order of magnitude to speed up the calculations, as in this energy regime the results are much less sensitive to the value of this parameter.

The calculations of $S(E)$ were carried out for $250$ values of energy distributed logarithmically within the range from $10^{-11}$ to $1$~a.u. 
For each energy, the absolute phase shifts were calculated until the relative contribution of a particular $l$ in Eq.~(\ref{eq:S_E}) was smaller than $10^{-8}$. 
If contributions from $l>16\,000$ were still significant, the Born approximation 
\begin{equation}\label{eq:wkb}
\mathrm{tan}\,\delta_l(E) \approx \frac{24\pi\mu_a^3C_6E^2}
{(2l-3)(2l-1)(2l+1)(2l+3)(2l+5)},
\end{equation}
was used to calculate the remaining terms necessary to reach the threshold. The quantity
$C_6$ in Eq.~(\ref{eq:wkb}) denotes the total asymptotic coefficient of the interaction potential, i.e., including the BO, relativistic, and QED contributions. 
The Born approximation was required for energies $E>10^{-3}$~a.u., where $S(E)$ is negative. 
For lower energies, angular momenta up to about $l=8\,000$ were sufficient.

The original grid of $250$ energies was not fine enough to properly describe the collision cross-sections $Q^{(n)}(E)$ as their behavior is less regular than that of $S(E)$.
Therefore, we created an additional set of $12\,000$ energies distributed non-uniformly according to
\begin{equation}
E_m/\,\mathrm{K} = 2\,(1.0005^m-1),\quad m\in\{1,\cdots,12\,000\}.
\end{equation}
For each energy, the summation in Eq.~(\ref{eq:Q_n}) was truncated when the relative contribution from a given $l$ was smaller than $10^{-12}$. 
The contributions to $Q^{(n)}(E)$ decay much faster with increasing $l$ than in the case of $S(E)$, so $l\approx 700$, on average, was sufficient to reach this threshold with the exception of higher energies for which $l\approx 1700$ was necessary. 

Calculated values of $S(E)$ and $Q^{(n)}(E)$, $n=2,4,6$, can be found in Supplemental Material \cite{supp}.

\subsection{Uncertainty estimation \label{subsec:methods_error}}

The final values of the pressure and acoustic virial coefficients were calculated using binding energies and the $S(E)$ function obtained by solving Eq.~(\ref{eq:schrod}) with $U(R)=V(R)$, where $V(R)$ is the total potential defined in Eq.~(\ref{eq:Vtot}).
The uncertainties of $B(T)$ and $\beta(T)$ were estimated as half of the absolute difference between the results calculated with $U(R)=V(R)\pm\sigma(R)$, where $\sigma(R)$ is the potential uncertainty function discussed in Sec.~\ref{subsec:fitsigma}.
This method of estimating the uncertainties will be further referred to as the ``$\pm\sigma$'' approach.

Calculation of the transport properties consists of two major steps.
In the first step, the collision cross-sections $Q^{(n)}(E)$ are obtained from the solutions to Eq.~(\ref{eq:schrod}).
Next, the $\Omega^{(n,s)}(T)$ integrals are generated from $Q^{(n)}(E)$ and transport properties are calculated using these integrals.
The final values of transport properties were calculated using $\Omega^{(n,s)}(T)$ generated from the collision cross-sections $Q^{(n)}_0(E)$ obtained using $U(R)=V(R)$ in Eq.~(\ref{eq:schrod}).
In the spirit of the ``$\pm\sigma$'' approach, the uncertainties of the transport properties may be estimated from the results of two separate calculations.
In the first calculations, $\Omega^{(n,s)}(T)$ are generated from the collision cross-sections $Q^{(n)}_{+}(E)$ obtained with $U(R)=V(R)+\sigma(R)$.
In the second, the procedure starts with the collision cross-sections $Q^{(n)}_{-}(E)$ obtained with $U(R)=V(R)-\sigma(R)$.
It is known, however, that this approach leads to artificial underestimation of the uncertainties \cite{garberoglio2023ab}. 
In this work, we propose an alternative approach to avoid the underestimation of uncertainties in calculation of transport properties.

\begin{figure}
\includegraphics[width=1\columnwidth]{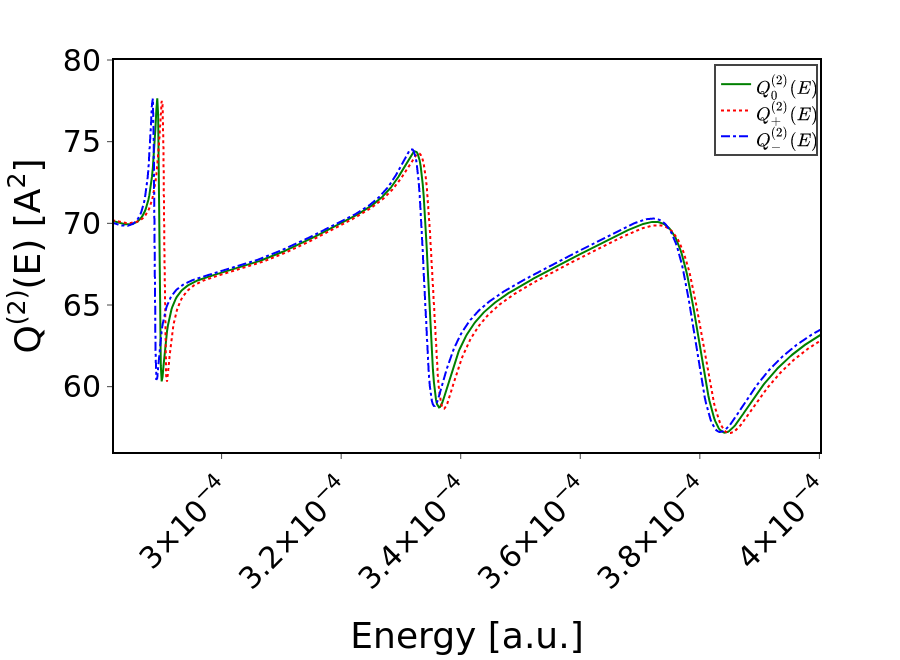}
\caption{Comparison of the quantum collision cross-section $Q^{(2)}_0(E)$ of the argon dimer calculated using the $V(R)$ potential and the collision cross-sections $Q^{(2)}_{\pm}(E)$ calculated using the perturbed potentials $V(R)\pm\sigma(R)$.
\label{fig:Qtest}}
\end{figure}

We found that the problem originates from the interplay between the behavior of the collision cross-sections as functions of the energy on the one side, and their dependence on the form of the potential $U(R)$ used in Eq.~(\ref{eq:schrod}) on the other side. 
In some energy regions the collision cross-sections $Q^{(n)}_0(E)$ are slightly smaller than $Q^{(n)}_{+}(E)$ but lie above $Q^{(n)}_{-}(E)$. 
This ordering is switched in other regions, see Fig.~\ref{fig:Qtest} for a representative example. 
The $\Omega^{(n,s)}(T)$ integrals defined in Eq.~(\ref{eq:omega}) are essentially weighted integrals of $Q^{(n)}(E)$ with the weight function $e^{-x}x^{s+1}\,dx$, where $x=E/k_\mathrm{B}T$. 
For any given temperature, this weight function is non-negligible on a large energy interval that includes several regions where $Q^{(n)}_{+}(E)$ and $Q^{(n)}_{-}(E)$ switch roles, i.e., the bound $Q^{(n)}_{-}(E)<Q^{(n)}_0(E)<Q^{(n)}_{+}(E)$ transforms into $Q^{(n)}_{+}(E)<Q^{(n)}_0(E)<Q^{(n)}_{-}(E)$ or the other way around.
When $Q^{(n)}_{+}(E)$, $Q^{(n)}_0(E)$, and $Q^{(n)}_{-}(E)$ are integrated to give the corresponding estimates of $\Omega^{(n,s)}(T)$, these oscillations between $Q^{(n)}_{+}(E)$ and $Q^{(n)}_{-}(E)$ artificially cancel out. 
This leads to unreasonably small differences between the values of $\Omega^{(n,s)}(T)$ obtained with $Q^{(n)}_{+}(E)$, $Q^{(n)}_0(E)$, and $Q^{(n)}_{-}(E)$, making the ``$\pm\sigma$'' approach unsuitable to estimate the uncertainty of $\Omega^{(n,s)}(T)$ and, finally, the transport properties.

To ameliorate this problem, one has to change the definition of the uncertainties of $Q^{(n)}(E)$ in the evaluation of $\Omega^{(n,s)}(T)$. We propose the following strategy comprising three steps:
\begin{itemize}
\item use $Q^{(n)}_\mathrm{max}(E) = \max (Q^{(n)}_0(E),Q^{(n)}_{+}(E),Q^{(n)}_{-}(E))$ in evaluation of Eq.~(\ref{eq:omega}) to obtain the upper bound of $\Omega^{(n,s)}(T)$;
\item use $Q^{(n)}_\mathrm{min}(E) = \min (Q^{(n)}_0(E),Q^{(n)}_{+}(E),Q^{(n)}_{-}(E))$ in evaluation of Eq.~(\ref{eq:omega}) to obtain the lower bound of $\Omega^{(n,s)}(T)$;
\item the uncertainties of the transport properties are estimated as half of the difference between the results obtained with the upper and lower bounds of $\Omega^{(n,s)}(T)$.
\end{itemize}
This strategy, further referred to as the ``$\max\!/\!\min Q$'' approach, leads to much more conservative estimates of the uncertainty and removes the problem of artificially small uncertainties for some temperatures. In Sec.~\ref{subsec:results_transport} we compare the uncertainties of the transport properties obtained with ``$\pm\sigma$'' and ``$\max\!/\!\min Q$'' approaches.

\section{Results and discussion \label{sec:results}}

\subsection{Spectroscopic parameters \label{subsec:results_spectro}}

\begin{table*}
\caption{Comparison of spectroscopic parameters and binding energies of $l=0$ vibrational states of the argon dimer supported by various theoretical potentials and determined from experiment. 
The position of the minimum $R_e$ is in angstrom and the depth of the potential well $D_e$, the rotational constant $B_0$, the centrifugal-distortion constant $D_0$, and the energy levels are in cm$^{-1}$. 
The values in parentheses are uncertainties of the rightmost digits.
\label{tab:vibration_levels}}
\begin{ruledtabular}
\begin{tabular}{cd{3.7}d{3.6}d{3.10}d{3.10}d{3.10}d{3.10}}
& 
\multicolumn{1}{c}{Patkowski 2010} & 
\multicolumn{1}{c}{Song 2019} & 
\multicolumn{3}{c}{This work, form of the potential $U(R)$ in Eq.~(\ref{eq:schrod})} &
\multicolumn{1}{c}{Expt.} \\
\cline{4-6}
& 
\multicolumn{1}{c}{Ref.~\cite{patkowski2010}} & 
\multicolumn{1}{c}{Ref.~\cite{song2019}} & 
\multicolumn{1}{c}{$V_\mathrm{BO}+V_\mathrm{CG}$} &
\multicolumn{1}{c}{$V_\mathrm{BO}+V_\mathrm{rel}+\delta V_\mathrm{ret}$} &
\multicolumn{1}{c}{$V$ in Eq.~(\ref{eq:Vtot})} &
\\
\hline
$R_e$ & 
3.762\footnote{\label{note:Pat}Taken from Ref.~\cite{patkowski2010}} & 
3.761 & 
3.7634(14) & 3.7639(14) & 3.7640(14) &
3.761(3)\footnote{\label{note:Her}Taken from Ref.~\cite{herman1988exp}}
\\
$D_e$ & 
-99.351\textsuperscript{\ref{note:Pat}} & 
-99.60 & 
-99.28(37) & -99.24(37) & -99.20(37) & 
-99.2(10)\textsuperscript{\ref{note:Her}} 
\\
$B_0$\footnote{\label{note:rot}Rotational and centrifugal-distortion constants were obtained from fitting of the lowest $v=0$ rotational states up to $l=14$.} &
0.057589\footnote{\label{note:Miz}Taken from Ref.~\cite{mizuse2022}} & 
0.057580 &
0.057560(45) & 0.057546(45) & 0.057543(46) & 
0.057611(4)\textsuperscript{\ref{note:Miz}} 
\\
$10^6\times D_0$\textsuperscript{\ref{note:rot}} &
1.03\textsuperscript{\ref{note:Miz}} &
0.99 & 
1.04(1) & 1.04(1) & 1.04(1) &
1.03(2)\textsuperscript{\ref{note:Miz}}
\\[0.5ex]
$v = 0$ & 
-84.53458\footnote{\label{note:Sah}Taken from Ref.~\cite{sahraeian2019}} & 
-84.40 & 
-84.50(34) & -84.46(34) & -84.43(34) & 
-84.47(1)\textsuperscript{\ref{note:Her}} 
\\
$v = 1$ & 
-58.85674\textsuperscript{\ref{note:Sah}} & 
-58.25 & 
-58.86(29) & -58.83(29) & -58.80(29) & 
-58.78(1)\textsuperscript{\ref{note:Her}} 
\\
$v = 2$ & 
-38.36106\textsuperscript{\ref{note:Sah}} & 
-37.61 & 
-38.38(24) & -38.35(24) & -38.33(24) & 
-38.20(2)\textsuperscript{\ref{note:Her}} 
\\
$v = 3$ & 
-22.85141\textsuperscript{\ref{note:Sah}} & 
-22.18 & 
-22.86(19) & -22.84(19) & -22.82(19) & 
-22.62(2)\textsuperscript{\ref{note:Her}} 
\\
$v = 4$ & 
-11.97942\textsuperscript{\ref{note:Sah}} & 
-11.51 & 
-11.98(13) & -11.96(13) & -11.95(13) & 
-11.71(3)\textsuperscript{\ref{note:Her}} 
\\
$v = 5$ &  
 -5.17193\textsuperscript{\ref{note:Sah}} &  
 -4.91 &  
 -5.17(8)  &  -5.15(8)  &  -5.15(8)  & 
 -4.87(7)\textsuperscript{\ref{note:Her}} 
\\
$v = 6$ &  
 -1.59539\textsuperscript{\ref{note:Sah}} &
 -1.48 & 
 -1.60(4)  &  -1.59(4)  &  -1.58(4)  &  
\multicolumn{1}{c}{-} 
\\
$v = 7$ &  
 -0.22722\textsuperscript{\ref{note:Sah}} & 
 -0.20 & 
 -0.23(1)  &  -0.22(1)  &  -0.22(1)  & 
\multicolumn{1}{c}{-} 
\\
$v = 8$ &
\multicolumn{1}{c}{$-0.20186\times10^{-6}$ \textsuperscript{\ref{note:Sah},}\footnote{From our calculations the energy is $-3.9\times10^{-6}$ cm$^{-1}$.}
 } &
\multicolumn{1}{c}{-} & 
\multicolumn{1}{c}{$[-2\times10^{-6}]$\footnote{\label{note:bracket}The potential $U(R)$ does not support the ninth vibrational state. The value in square brackets was obtained using $U(R)-\sigma(R)$, where $\sigma(R)$ is the uncertainty function.}} &
\multicolumn{1}{c}{$[-6\times10^{-6}]$\textsuperscript{\ref{note:bracket}}} &
\multicolumn{1}{c}{$[-6\times10^{-6}]$\textsuperscript{\ref{note:bracket}}} &
\multicolumn{1}{c}{-}
\\
\end{tabular}
\end{ruledtabular}
\end{table*}

In Table~\ref{tab:vibration_levels} we collect spectroscopic parameters of the argon dimer calculated using our potential. 
The results were obtained at three levels of theory, differing in the treatment of the post-BO effects:
\begin{itemize}
    \item using only the Cowan--Griffin one-electron relativistic correction;
    \item using the complete Breit--Pauli Hamiltonian including the two-electron terms;
    \item same as the above with addition of the QED correction.
\end{itemize}
The retardation effects were included at the last two levels and the uncertainties of the calculated quantities were obtained in all three cases using the uncertainty function $\sigma(R)$ discussed in Sec.~\ref{subsec:fitsigma}.  
For comparison, we also provide corresponding results for two potentials taken from the literature -- the \emph{ab initio} potential of Patkowski and Szalewicz \cite{patkowski2010} and the empirical potential of Song and Yang \cite{song2019}. The experimental results \cite{herman1988exp,mizuse2022} are also given. The energies of all rovibrational states supported by our total potential can be found in Supplemental Material \cite{supp}.

The theoretical potentials considered in Table~\ref{tab:vibration_levels} agree on the location of the minimum of the potential well at around $R=3.76$~\AA{} and on its depth in the range between $-99.6$ and $-99.2$~cm$^{-1}$. 
All results are within error bounds of the experimental determination from Ref.~\cite{herman1988exp}.
By contrast, theoretical rotational constants are outside very narrow error bounds of the experimental value from Ref.~\cite{mizuse2022} but the relative differences do not exceed 0.12\%. 
Experimental error bars of the centrifugal-distortion constant are much wider and only empirical potential of Song and Yang \cite{song2019} fails to reproduce the experimental value.
The energies of rotationless vibrational levels supported by each potential are close to available experimental data. 
Comparing the results obtained at different levels of accounting for the post-BO effects, it is seen that inclusion of the two-electron relativistic terms and retardation effects improves the agreement with the experiment.
When the QED correction is added, the energies predicted using our potential are the closest to the experimental values for $v=1\text{-}3$.

We obtained at least eight bound vibrational states with each potential, while the ninth bound state was observed only with the potential of Patkowski and Szalewicz \cite{patkowski2010}.
Nevertheless, based on the estimated uncertainty of our data, we cannot rule out the possibility that the ninth bound state exists. It appears when our potential is modified by subtracting the uncertainty function $\sigma(R)$, i.e., when the estimated lower-bound of our potential is used.
Moreover, for the ninth state to appear, it is sufficient to subtract $0.3\text{-}0.4\sigma(R)$ when the description of the post-BO effects is restricted to the Cowan--Griffin correction alone, or $0.7\text{-}0.8\sigma(R)$ when the total potential is used. 
These observations suggest that in order to conclusively answer the question of how many rotationless vibrational levels are supported by the electronic ground state of the argon dimer, it is necessary to reduce the uncertainty of the \emph{ab initio} data by a factor of at least two. 
However, the knowledge of the existence and precise position of the ninth vibrational state becomes important in the calculations of thermophysical properties of the argon gas only for temperatures well below 1~K, and is therefore irrelevant to the results discussed in the following sections.

\subsection{Second pressure and acoustic virial coefficients \label{subsec:results_virial}}

\begin{table*}
\caption{The results (in $\mathrm{cm}^3\,\mathrm{mol}^{-1}$) obtained in this work from quantum-mechanical calculations of the second pressure virial coefficient $B_\mathrm{qm}(T)$ and the second acoustic virial coefficient $\beta_\mathrm{qm}(T)$. 
The values in parentheses represent uncertainties of the rightmost digits. 
Also shown are relative deviations of the semiclassical results taken from Refs.~\cite{vogel2010pot} and \cite{moldover2014acoustic}, and of the semiclassical results obtained using the potential from Ref.~\cite{deiters2019two} and the potential developed in this work, from the quantum-mechanical values [$\Delta B_\mathrm{scl}=(B_\mathrm{scl}-B_\mathrm{qm})/B_\mathrm{qm}\times100\%$ and $\Delta \beta_\mathrm{scl}=(\beta_\mathrm{scl}-\beta_\mathrm{qm})/\beta_\mathrm{qm}\times100\%$].
\label{tab:secondcompar}}
\begin{ruledtabular}
\begin{tabular}{d{3.0}d{4.6}d{2.4}d{2.4}d{2.4}d{2.4}d{4.6}d{2.4}d{2.4}d{2.4}d{2.4}}
& 
\multicolumn{1}{c}{$B_\mathrm{qm}(T)$} &
\multicolumn{4}{c}{$\Delta B_\mathrm{scl}(T)$} & 
\multicolumn{1}{c}{$\beta_\mathrm{qm}(T) $}  &
\multicolumn{4}{c}{$\Delta \beta_\mathrm{scl}(T)$}
\\ 
\cline{3-6} \cline{8-11}
\multicolumn{1}{c}{$T$ [K]} & 
& 
\multicolumn{1}{c}{Ref.~\cite{vogel2010pot}} & 
\multicolumn{1}{c}{Ref.~\cite{moldover2014acoustic}} &  
\multicolumn{1}{c}{Ref.~\cite{deiters2019two}} &  
\multicolumn{1}{c}{this work} & 
& 
\multicolumn{1}{c}{Ref.~\cite{vogel2010pot}} & 
\multicolumn{1}{c}{Ref.~\cite{moldover2014acoustic}} &
\multicolumn{1}{c}{Ref.~\cite{deiters2019two}} &  
\multicolumn{1}{c}{this work} 
\\ 
\hline
90  & -220.9(16) &  0.35\,\% &  0.18\,\% &  0.53\,\% &  0.01\,\% & -228.3(20)   &  0.45\,\% &  0.18\,\% &  0.67\,\% & <0.01\,\% \\
110 & -152.6(11) &  0.36\,\% &  0.18\,\% &  0.55\,\% & <0.01\,\% & -144.3(13)   &  0.46\,\% &  0.20\,\% &  0.72\,\% & <0.01\,\% \\
200 & -47.68(43) &  0.46\,\% &  0.23\,\% &  0.76\,\% & <0.01\,\% &  -23.69(53)  &  1.07\,\% &  0.49\,\% &  1.78\,\% & -0.01\,\% \\
280 & -19.39(29) &  0.73\,\% &  0.35\,\% &  1.24\,\% & -0.01\,\% &    7.41(36)  & -2.16\,\% & -0.91\,\% & -3.60\,\% & <0.01\,\% \\
340 &  -8.15(24) &  1.36\,\% &  0.64\,\% &  2.34\,\% & -0.01\,\% &    19.49(30) & -0.63\,\% & -0.24\,\% & -1.04\,\% & <0.01\,\% \\
500 &   7.08(16) & -0.98\,\% & -0.40\,\% & -1.66\,\% &  0.01\,\% &    35.31(21) & -0.21\,\% & -0.05\,\% & -0.31\,\% & <0.01\,\% \\
\end{tabular}
\end{ruledtabular}
\end{table*}

To the best of our knowledge, in this work we report the first fully quantum-mechanical calculations of the second virial coefficients of argon gas with rigorous error estimates. 
The results for a selected set of temperatures are presented in Table~\ref{tab:secondcompar}.
The relative uncertainties of the calculated second virial coefficients are below $1.3\%$ for all considered temperatures, except for the regions where the sign change occurs -- at about $408$~K for the second pressure virial coefficient and $255$~K for the second acoustic virial coefficient. 
The full set of calculated second pressure and acoustic virial coefficients can be found in Supplemental Material \cite{supp}.

\begin{figure*}
\begin{subfigure}{.8\linewidth}
  \includegraphics[width=1\linewidth]{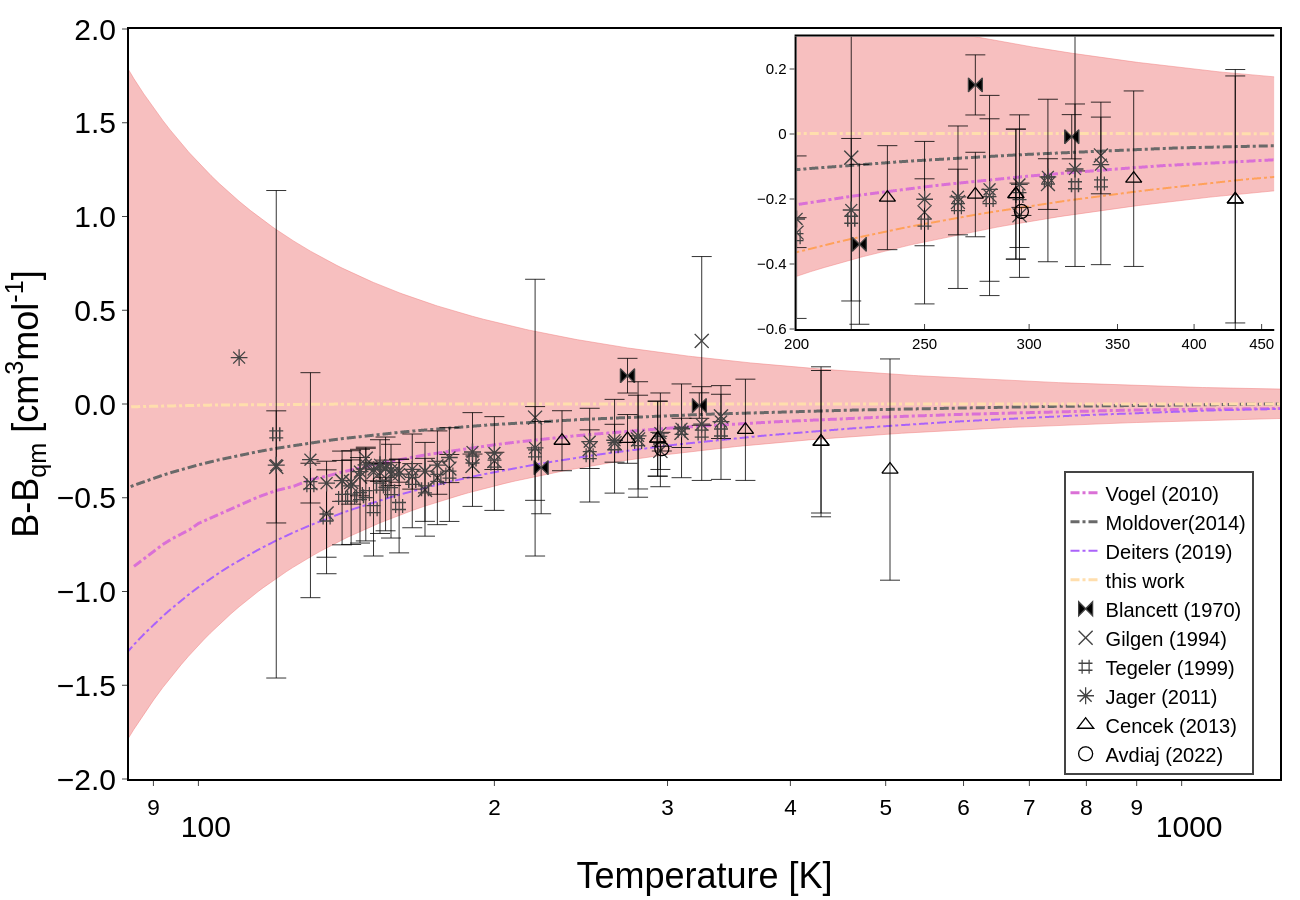} 
\end{subfigure}
\begin{subfigure}{.8\linewidth}
  \includegraphics[width=1\linewidth]{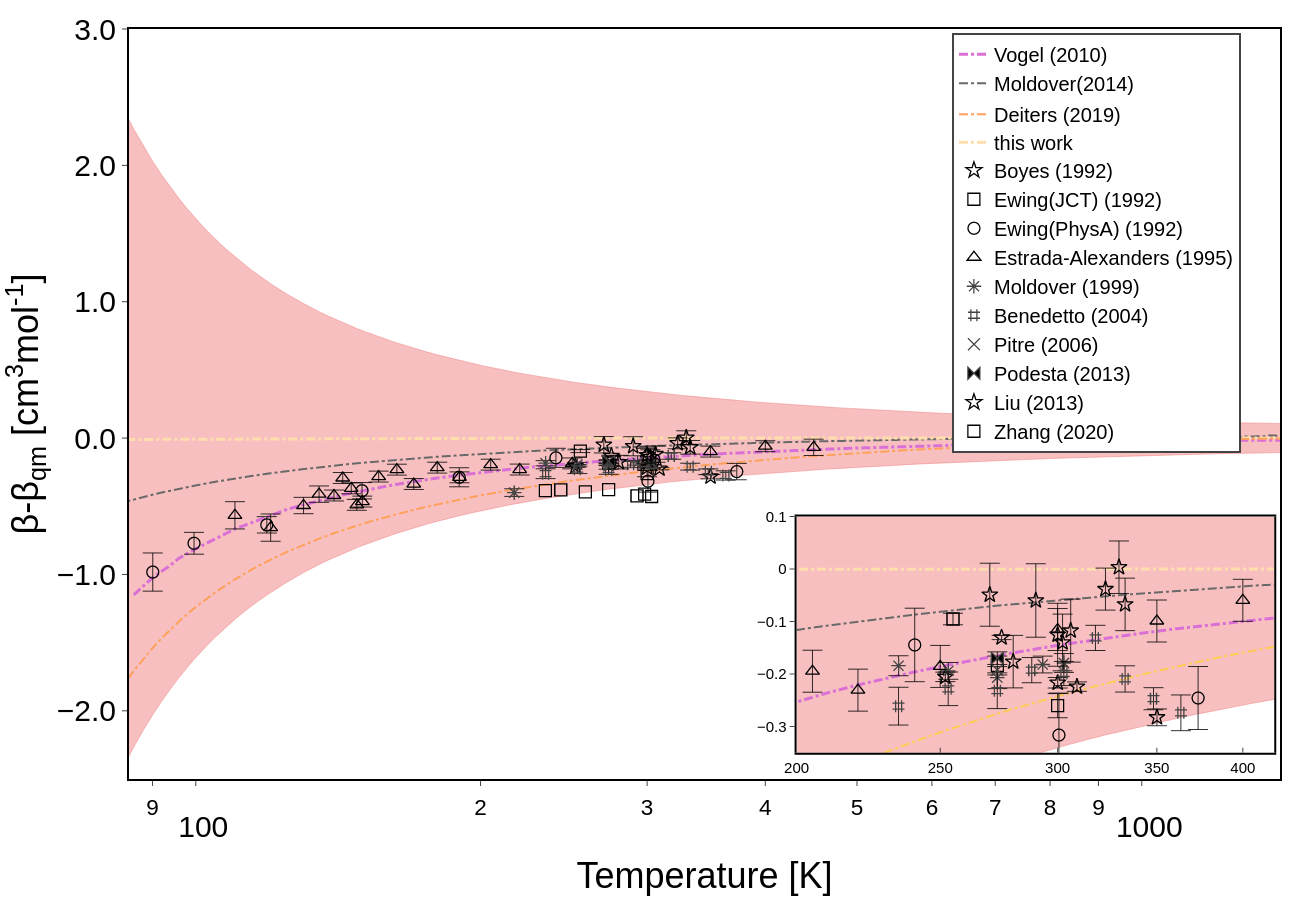}
\end{subfigure}
\caption{Comparison of the quantum-mechanical values of the second pressure virial coefficient $B_\mathrm{qm}(T)$ (upper panel) and the second acoustic virial coefficient $\beta_\mathrm{qm}(T)$ (lower panel) obtained in this work with selected theoretical and experimental results.
Dash-dotted lines correspond to the semiclassical results taken from Refs.~\cite{vogel2010pot} and \cite{moldover2014acoustic} and the semiclassical values obtained using the potential from Ref.~\cite{deiters2019two} and the potential developed in this work. 
Points with error bars correspond to the experimental results taken from Refs.~\cite{blancett1970,boyes1992,EWING1992531,EWING1992415,GILGEN1994383,ESTRADAALEXANDERS19951075,moldover1999,tegeler1999new,benedetto2004,pitre2006acoustic,cencek2013,podesta2013low,liu2013determination,zhang2020determination,avdiaj2022measurements}. 
Red area represents the estimated uncertainty of our data.
\label{fig:second}}
\end{figure*}

In Table~\ref{tab:secondcompar} and Fig.~\ref{fig:second} we compare the results of our quantum-mechanical calculations with the values obtained using semiclassical approaches, including the results of Vogel \emph{et al.}\ \cite{vogel2010pot} calculated using the potential of J\"ager \emph{et al.}\ \cite{jager2009}, the results of Moldover \emph{et al.}\ \cite{moldover2014acoustic} based on the potential of Patkowski and Szalewicz \cite{patkowski2010}, and semiclassical results calculated using the quadratic Feynman--Hibbs method \cite{feynman1965pathintegrals}  with the potential of Deiters and Sadus \cite{deiters2019two} and the potential developed in this work.
In Fig.~\ref{fig:second} we also compare our data with selected experimental values.
Recently, Myatt \emph{et al.}\ \cite{myatt2018} have consolidated and reevaluated available experimental results for several thermophysical properties of argon gas \cite{johnston1942,MICHELS1949627,flynn1963,kestin1964,clarke1968,gracki1969,guevara1969,dawe1970,kestin1972,faubert1972,haarman1973,springer1973,hellemans1974,chen1975,clifford1975,clifford1981,FLEETER1981371,ziebland1981,kestin1982,haran1983,kestin1984,vargaftik1984,johns1986,mardolcar1986,younglove1986,MILLAT1987461,hemminger1987,moldover1988,EWING1989899,perkins1991,EWING1992415,EWING1992531,GILGEN1994383,ESTRADAALEXANDERS19951075,estradaalexanders1996,sevastyanov1996,KLIMECK19981571,moldover1999,roder2000,wilhelm2000,evers2002,benedetto2004,sun2005,may2007,vogel2010,berg2012,berg2013,cencek2013,lin2014}. 
As older experiments exhibit large uncertainties, we omit them in Fig.~\ref{fig:second} and focus on the most recent experimental determinations with more narrow error bars \cite{blancett1970,boyes1992,EWING1992531,EWING1992415,GILGEN1994383,ESTRADAALEXANDERS19951075,moldover1999,tegeler1999new,benedetto2004,pitre2006acoustic,cencek2013,podesta2013low,liu2013determination,zhang2020determination,avdiaj2022measurements}.

The experimental results are within estimated uncertainties of our quantum-mechanical calculations.
It is worth to note that the experimental error bars are generally narrower than our uncertainty estimates -- by a factor of 2-3 for the second pressure virial coefficient and by almost an order of magnitude in the case of the second acoustic virial coefficient.
The main source of uncertainty in our calculations is the BO contribution to the pair potential taken from Refs.~\cite{patkowski2005,patkowski2010}, as other sources of error, resulting from the treatment of relativistic and QED effects used in this work, are negligible compared to the estimated theoretical uncertainties of BO energies.

Interestingly, the second acoustic virial coefficients obtained by Vogel \emph{et al.}\ \cite{vogel2010pot} from the potential of J\"ager \emph{et al.}\ \cite{jager2009} are almost identical to the experimental values. 
This is somewhat surprising as J\"ager \emph{et al.}\ \cite{jager2009} performed their \emph{ab initio} calculations at the same level of theory as Patkowski and Szalewicz \cite{patkowski2010} but employed smaller basis sets. 
Not surprisingly, the results of Moldover \emph{et al.}\ are closer to ours as we share the same BO interaction energies from Ref.~\cite{patkowski2010}.
While Patkowski \emph{et al.}\ \cite{patkowski2005,patkowski2010} provided uncertainty estimates for their \emph{ab initio} energies at a discrete set of internuclear distances, no systematic uncertainty estimation of the analytic potential developed in Ref.~\cite{patkowski2010} was given. 
Therefore, it is unknown how Moldover \emph{et al.}\ \cite{moldover2014acoustic} obtained their estimations of uncertainty, but their uncertainties generally agree with ours obtained using the ``$\pm\sigma$'' approach.  
The results calculated from the simplified potential of Deiters and Sadus
\cite{deiters2019two} that does not include any quantum contributions differ the most
from our full quantum-mechanical values.
Overall, all analyzed theoretical potentials lead to results that are within 2-3\% of the experimental data within the temperature range where the second virial or acoustic coefficient change sign. For the remaining temperatures, the differences are well within 1\%.
Therefore, all theoretical potentials can be seen as quantitatively comparable. 
Nevertheless, only the potential developed in this work properly includes all relevant physical effects, some of which have been neglected in previous studies. We also provide an analytical fit of the uncertainty of the interaction potential which has not been reported in other theoretical works.

The semiclassical results obtained using our potential are almost equal to the quantum-mechanical ones.
This suggests that the semiclassical calculations are sufficient for the accurate determination of the second pressure and acoustic virial coefficients of argon. 
Indeed, the relative deviations between the semiclassical and quantum-mechanical calculations are $\le0.02\%$ within the studied temperature range (30-4000~K). 
The deviations increase slightly for smaller temperatures but even at 30~K, i.e., way below the freezing point of argon (83.95~K), the deviation is $\approx$ 0.02\%. 
We expect the same behavior for heavier noble gases such as krypton and xenon as their freezing points are much higher. 

Using the semiclassical approach we also calculated the second virial coefficients of argon using other recent theoretical potentials from the literature \cite{myatt2018,song2019,sheng2020conformal} not shown in Table~\ref{tab:secondcompar} and Fig.~\ref{fig:second}.
The values calculated using the potential of Myatt \emph{et al.}\ \cite{myatt2018} exhibits small deviations from our results and are well within estimated uncertainties of this work for most of the studied temperatures.
By contrast, the values obtained using the potentials of Song and Yang \cite{song2019} and Sheng \emph{et al.}\ \cite{sheng2020conformal} deviate from our results by as much as 6-10\% and 4-10\%, respectively, i.e., they are well outside both our estimated uncertainties and the error bars of the newest experimental determinations. 
Therefore, although the latter potentials are able to accurately predict vibrational excitation energies of the argon dimer \cite{sheng2020conformal}, they are not adequate for the calculations of the second pressure and acoustic virial coefficients. 

\subsection{Transport properties \label{subsec:results_transport}}

\begin{figure}
\includegraphics[width=1\columnwidth]{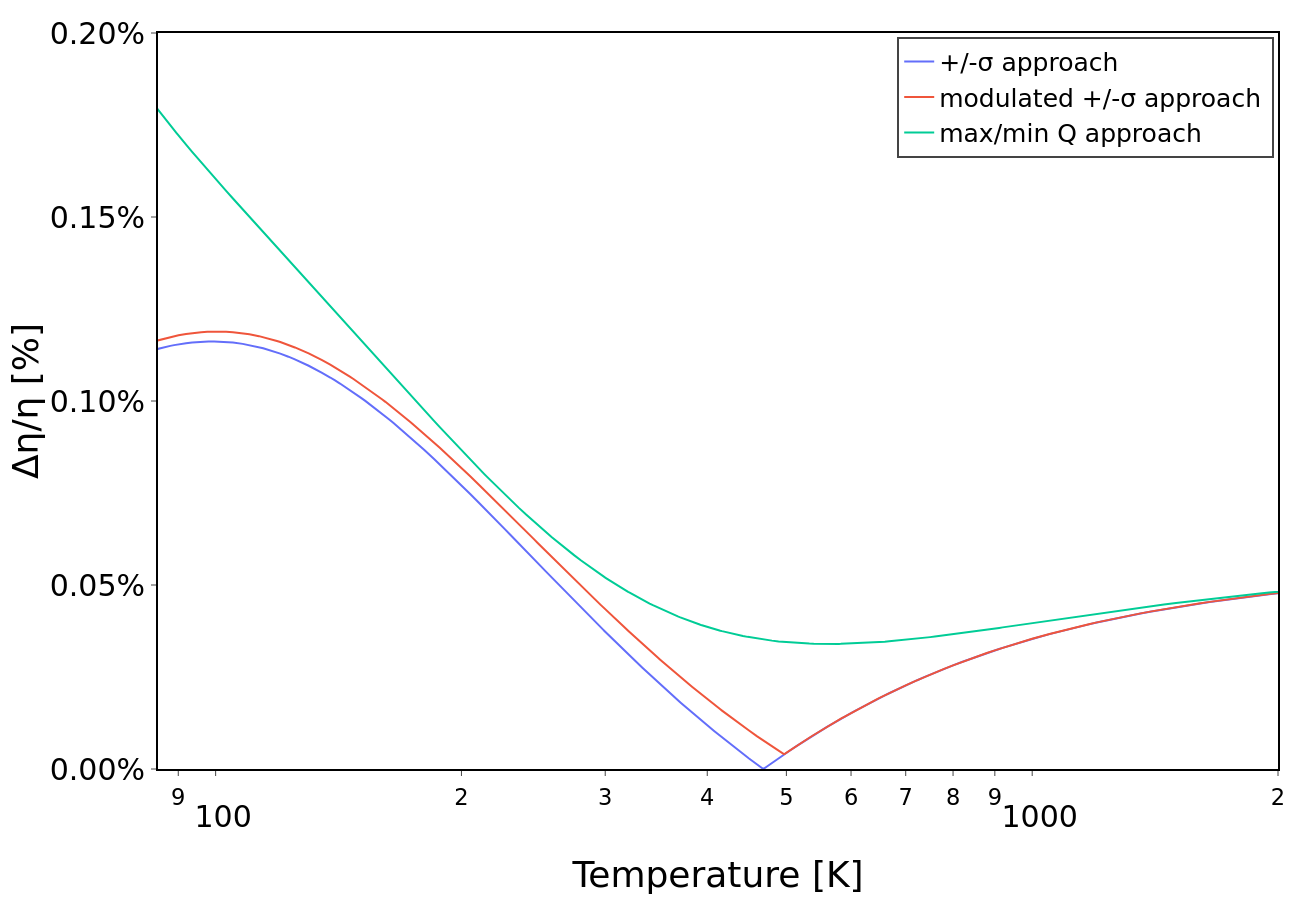}
\caption{Relative uncertainties of the viscosity estimated using various approaches. \label{fig:transport_unc}}
\end{figure}

As discussed in Sec.~\ref{subsec:methods_error}, a rigorous and reliable estimation of the uncertainty of theoretical transport properties is a challenging task.
In Fig.~\ref{fig:transport_unc}, we present the uncertainties of the viscosity calculated with our potential using three different methods: the simple ``$\pm\sigma$'' approach, the modulated ``$\pm\sigma$'' approach proposed by Hellmann and co-workers \cite{hellmann2016Kr,hellmann2017Xe,hellmann2021thermophysicalNeon}, and the ``$\max\!/\!\min Q$'' approach suggested in this work.
The uncertainties calculated using the ``$\pm\sigma$'' approach become close to zero for temperatures around 450~K which is an artifact.
Analogous behavior is also observed in the theoretical results of Moldover \emph{et al.}\ \cite{moldover2014acoustic} at a similar temperature. 
The modulated ``$\pm\sigma$'' approach partially fixes this problem, but the uncertainties are still underestimated in the vicinity of the artificial minimum of the curve.
By contrast, in the ``$\max\!/\!\min Q$'' approach only a very shallow minimum is present.
Moreover, while for high temperatures the ``$\max\!/\!\min Q$'' method provides almost identical results as both the simple and modulated ``$\pm\sigma$'' methods, the uncertainties for low temperatures are about 1.6 times larger. 
This suggests that the latter methods may also underestimate the uncertainties for low temperatures.

The full set of calculated thermal conductivities and viscosities, together with their estimated uncertainties, can be found in Supplemental Material \cite{supp}.
In Fig.~\ref{fig:transport} we compare the results of our quantum-mechanical calculations with the existing theoretical and experimental determinations.
The theoretical values include the results from classical calculations of Vogel \emph{et al.}\ \cite{vogel2010pot} and Moldover \emph{et al.}\ \cite{moldover2014acoustic}, and quantum-mechanical calculations of Sharipov and Benites \cite{sharipov2019transport} based on the potential of Patkowski and Szalewicz \cite{patkowski2010}. 
Sharipov and Benites \cite{sharipov2019transport} also provided estimations of the uncertainty, but only as a difference between the values obtained with the potentials from Refs.~\cite{jager2009} and \cite{patkowski2010}. 
Somewhat surprisingly, their estimated uncertainties for temperatures below 300~K are smaller than the differences between the values of Vogel \emph{et al.}\ \cite{vogel2010pot} and Moldover \emph{et al.}\ \cite{moldover2014acoustic} that were calculated using the same potentials from Refs.~\cite{jager2009} and \cite{patkowski2010}, respectively.
The results from the classical calculations deviate significantly from ours, and for temperatures $>150$~K they are outside our uncertainty estimates.
By contrast, the results of Sharipov and Benites \cite{sharipov2019transport} for the thermal conductivity are similar to our results for temperatures below 400~K, but differ by about 0.07\% for higher temperatures. 
In the case of the viscosity, the deviations are slightly larger but mostly within the estimated uncertainties of our calculations. 

A large number of experimental work on the viscosity and thermal conductivity has been conducted on argon gas \cite{johnston1942,flynn1963,kestin1964,clarke1968,gracki1969,guevara1969,dawe1970,kestin1972,faubert1972,haarman1973,springer1973,hellemans1974,clifford1975,chen1975,FLEETER1981371,clifford1981,ziebland1981,kestin1982,haran1983,kestin1984,vargaftik1984,johns1986,mardolcar1986,younglove1986,MILLAT1987461,hemminger1987,perkins1991,roder2000,wilhelm2000,evers2002,sun2005,may2007,vogel2010,berg2012,berg2013,zhang2013effects,lin2014,xiao2020wide}. 
In general, the experimental values of argon transport properties come from relative measurements, where argon is measured relative to helium for which highly accurate \emph{ab initio} values are known.
The argon/helium viscosity ratio is typically measured directly.
Thermal conductivity, on the other hand, is usually obtained from viscosity using \emph{ab initio} calculations of the Prandtl number since measurements of vapor thermal conductivity have larger uncertainties.
Similarly as in the case of virial coefficients, in Fig.~\ref{fig:transport} we show only the experimental data that have been reported recently and have small uncertainties 
\cite{sun2005,may2007,vogel2010,berg2012,berg2013,zhang2013effects,lin2014,xiao2020wide}. 
In contrast to the second virial coefficients, the experimental uncertainties for the transport properties exceed the theoretical uncertainties reported in this work. 
Nevertheless, the results are consistent within combined error bounds.

\begin{figure*}
\begin{subfigure}{0.8\linewidth}
  \includegraphics[width=1\linewidth]{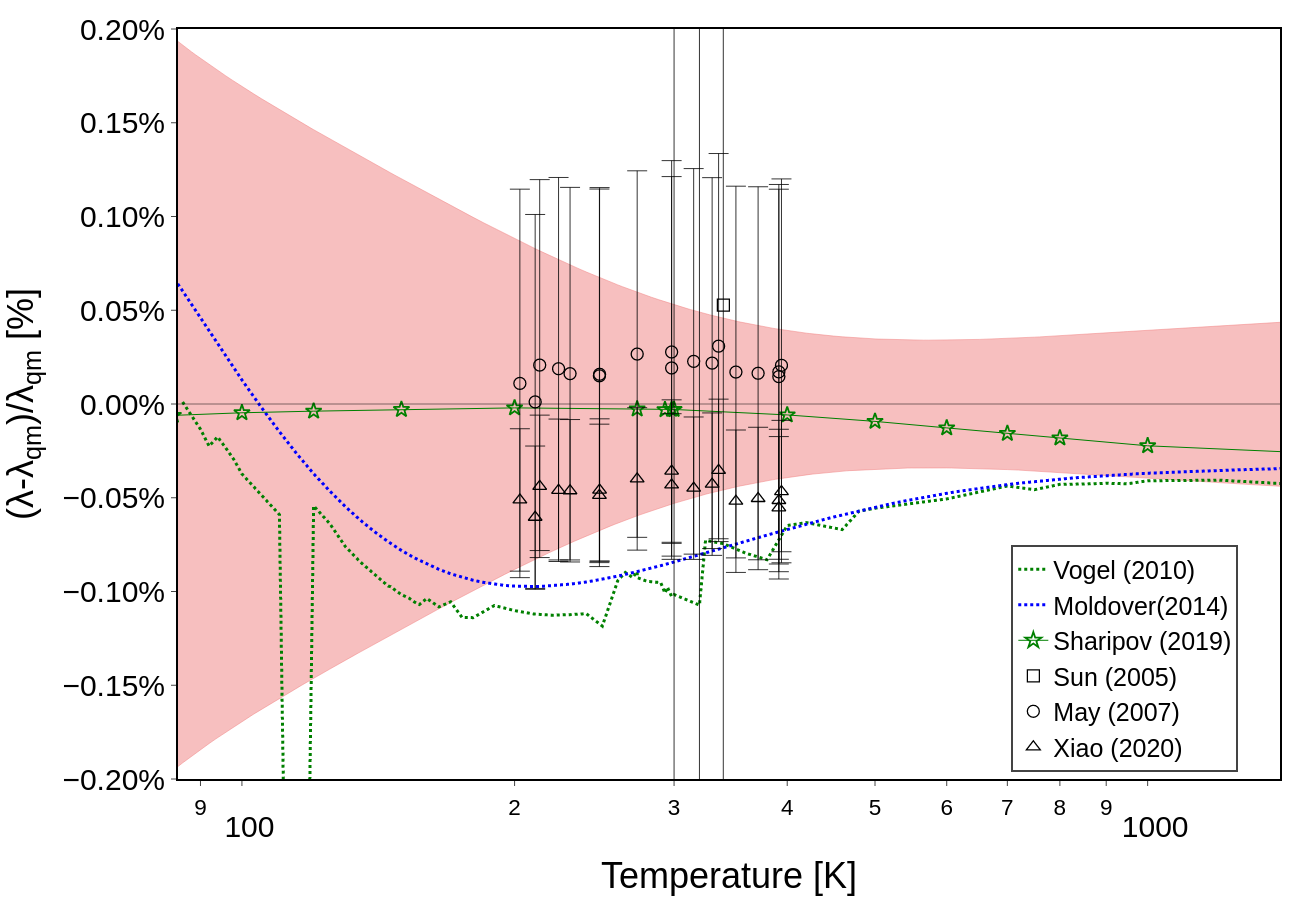}
\end{subfigure}
\begin{subfigure}{.8\linewidth}
  \includegraphics[width=1\linewidth]{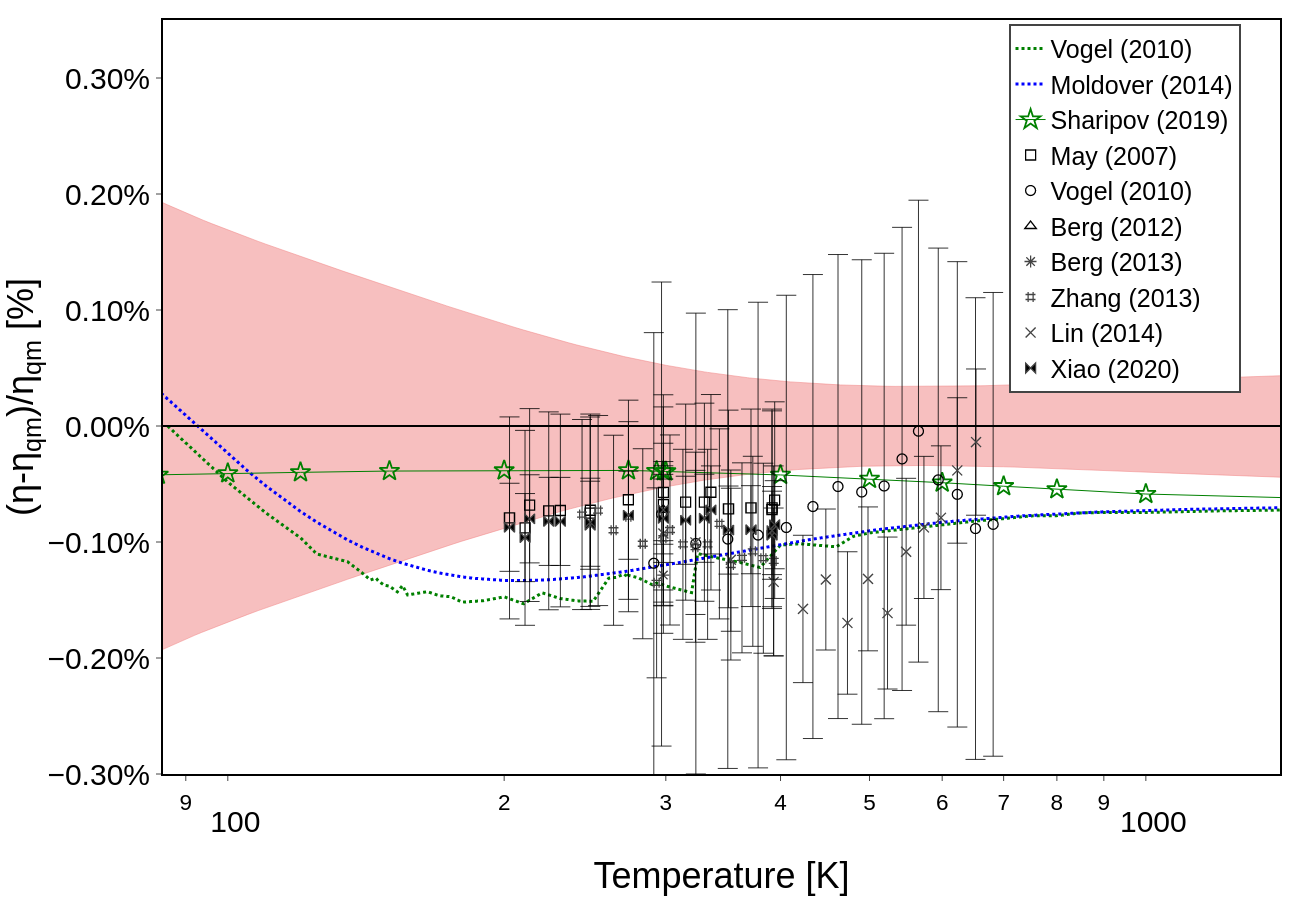}
\end{subfigure}
\caption{Comparison of the quantum-mechanical values of the thermal conductivity $\lambda_\mathrm{qm}(T)$ (upper panel) and the viscosity $\eta_\mathrm{qm}(T)$ (lower panel) obtained in this work with selected theoretical and experimental results.
Dotted lines correspond to the classical results taken from Refs.~\cite{vogel2010pot} and \cite{moldover2014acoustic} and solid line with stars represents quantum-mechanical calculations from Ref.~\cite{sharipov2019transport}.
Points with error bars correspond to the experimental results taken from Refs.~\cite{sun2005,may2007,vogel2010,berg2012,berg2013,zhang2013effects,lin2014,xiao2020wide}.
Red area represents the estimated uncertainty of our data.
\label{fig:transport}}
\end{figure*}

\section{Conclusions \label{sec:conclusion}}

In this work, we have developed an \emph{ab initio} interaction potential for the electronic ground state of the argon dimer. 
The Born--Oppenheimer component of the potential has been taken from Refs.~\cite{patkowski2005,patkowski2010}. 
However, we have refined the potential by including all sizeable post-Born--Oppenheimer effects. 
First, the relativistic corrections have been calculated using the full Breit--Pauli Hamiltonian, taking into account both the one- and two-electron effects. 
Second, the leading-order quantum electrodynamics correction has been calculated and the retardation of the electromagnetic interactions has been included in the long-range part of the potential. 
Finally, we provide an analytical fit of the uncertainties of the calculated contributions to the interaction potential which is not available in other theoretical studies.

Spectroscopic properties of the argon dimer such as rovibrational energy levels, bond-dissociation energy, and rotational and centrifugal-distortion constants have been reported. 
We have shown that at the current accuracy level it is not possible to determine whether the weakly bound ninth rotationless vibrational level exists or not.
Thermophysical properties of the argon gas -- pressure and acoustic virial coefficients, as well as transport properties -- viscosity and thermal conductivity -- have been determined using the developed potential. 
In the case of the thermophysical properties, the theoretical values reported here are somewhat less accurate than the most recent experimental data.
However, the opposite is true for the transport properties -- theoretical results calculated in this work have considerably smaller uncertainties than the data derived from measurements.

\begin{acknowledgments}
We thank Allan H. Harvey and Bogumi{\l} Jeziorski for useful comments on the manuscript, and Pawe{\l} Czachorowski for making the dataset of phase shifts used in Ref.~\cite{czachorowski2020} available to us. We acknowledge support from the Real-K project 18SIB02, which has received funding from the EMPIR programme cofinanced by the Participating States and from the European Union’s Horizon 2020 research and innovation programme. The support from the National Science Center, Poland, within Project No.\ 2017/27/B/ST4/02739 is also acknowledged. We gratefully acknowledge Poland's high-performance Infrastructure PLGrid (HPC Centers: ACK Cyfronet AGH, PCSS, CI TASK, WCSS) for providing computer facilities and support within computational grants no. PLG/2023/016440 and PLG/2023/016599.
\end{acknowledgments}

\bibliography{argon2}

\end{document}